**A ring-like accretion structure in M87 connecting its black hole and jet**

Ru-Sen Lu[1,2,3,*], Keiichi Asada[4,*], Thomas P. Krichbaum[3,*], Jongho Park[4,5], Fumie Tazaki[6,7], Hung-Yi Pu[8,4,9], Masanori Nakamura[10,4], Andrei Lobanov[3], Kazuhiro Hada[7,11,*], Kazunori Akiyama[12,13,14], Jae-Young Kim[15,5,3], Ivan Marti-Vidal[16,17], José L. Gómez[18], Tomohisa Kawashima[19], Feng Yuan[1,20,21], Eduardo Ros[3], Walter Alef[3], Silke Britzen[3], Michael Bremer[22], Avery E. Broderick[23,24,25], Akihiro Doi[26,27], Gabriele Giovannini[28,29], Marcello Giroletti[29], Paul T. P. Ho[4], Mareki Honma[7,11,30], David H. Hughes[31], Makoto Inoue[4], Wu Jiang[1], Motoki Kino[14,32], Shoko Koyama[33,4], Michael Lindqvist[34], Jun Liu[3], Alan P. Marscher[35], Satoki Matsushita[4], Hiroshi Nagai[14,11,] Helge Rottmann[3], Tuomas Savolainen[36,37,3], Karl-Friedrich Schuster[22], Zhi-Qiang Shen[1,2], Pablo de Vicente[38], R. Craig Walker[39], Hai Yang[1,21], J. Anton Zensus[3], Juan Carlos Algaba[40], Alexander Allardi[41], Uwe Bach[3], Ryan Berthold[42], Dan Bintley[42], Do-Young Byun[5,43], Carolina Casadio[44,45], Shu-Hao Chang[4], Chih-Cheng Chang[46], Song-Chu Chang[46], Chung-Chen Chen[4], Ming-Tang Chen[47], Ryan Chilson[47], Tim C. Chuter[42], John Conway[34], Geoffrey B. Crew[13], Jessica T. Dempsey[42,48], Sven Dornbusch[3], Aaron Faber[49], Per Friberg[42], Javier González García[38], Miguel Gómez Garrido[38], Chih-Chiang Han[4], Kuo-Chang Han[46], Yutaka Hasegawa[50], Ruben Herrero-Illana[51], Yau-De Huang[4], Chih-Wei L. Huang[4], Violette Impellizzeri[52,53], Homin Jiang[4], Hao Jinchi[54], Taehyun Jung[5], Juha Kallunki[37], Petri Kirves[37], Kimihiro Kimura[55], Jun Yi Koay[4], Patrick M. Koch[4], Carsten Kramer[22], Alex Kraus[3], Derek Kubo[47], Cheng-Yu Kuo[56], Chao-Te Li[4], Lupin Chun-Che Lin[57], Ching-Tang Liu[4], Kuan-Yu Liu[4], Wen-Ping Lo[58,4], Li-Ming Lu[46], Nicholas MacDonald[3], Pierre Martin-Cocher[4], Hugo Messias[59,51], Zheng Meyer-Zhao[48,4], Anthony Minter[60], Dhanya G. Nair[61], Hiroaki Nishioka[4], Timothy J. Norton[62], George Nystrom[47], Hideo Ogawa[50], Peter Oshiro[47], Nimesh A Patel[62], Ue-Li Pen[4], Yurii Pidopryhora[3,63], Nicolas Pradel[4], Philippe A. Raffin[47], Ramprasad Rao[62], Ignacio Ruiz[64], Salvador Sanchez[64], Paul Shaw[4], William Snow[47], T. K. Sridharan[53,62], Ranjani Srinivasan[62,4], Belén Tercero[38], Pablo Torne[64], Efthalia Traianou[18,3], Jan Wagner[3], Craig Walther[42], Ta-Shun Wei[4], Jun Yang[34], Chen-Yu Yu[4]

1 Shanghai Astronomical Observatory, Chinese Academy of Sciences, Shanghai, People's Republic of China

2 Key Laboratory of Radio Astronomy, Chinese Academy of Sciences, Nanjing, People's Republic of China

3 Max-Planck-Institut für Radioastronomie, Bonn, Germany

4 Institute of Astronomy and Astrophysics, Academia Sinica, Taipei, Taiwan, ROC
5 Korea Astronomy and Space Science Institute, Daejeon, Republic of Korea
6 Simulation Technology Development Department, Tokyo Electron Technology Solutions, Oshu, Japan
7 Mizusawa VLBI Observatory, National Astronomical Observatory of Japan, Oshu, Japan
8 Department of Physics, National Taiwan Normal University, Taipei, Taiwan, ROC
9 Center of Astronomy and Gravitation, National Taiwan Normal University, Taipei, Taiwan, ROC
10 Department of General Science and Education, National Institute of Technology, Hachinohe College, Hachinohe City, Japan
11 Department of Astronomical Science, The Graduate University for Advanced Studies, SOKENDAI, Mitaka, Japan
12 Black Hole Initiative, Harvard University, Cambridge, MA, USA
13 Massachusetts Institute of Technology Haystack Observatory, Westford, MA, USA
14 National Astronomical Observatory of Japan, Mitaka, Japan
15 Department of Astronomy and Atmospheric Sciences, Kyungpook National University, Daegu, Republic of Korea
16 Departament d'Astronomia i Astrofísica, Universitat de València, Valencia, Spain
17 Observatori Astronòmic, Universitat de València, Valencia, Spain
18 Instituto de Astrofísica de Andalucía-CSIC, Granada, Spain
19 Institute for Cosmic Ray Research, The University of Tokyo, Chiba, Japan
20 Key Laboratory for Research in Galaxies and Cosmology, Chinese Academy of Sciences, Shanghai, People's Republic of China
21 School of Astronomy and Space Sciences, University of Chinese Academy of Sciences, Beijing, People's Republic of China
22 Institut de Radioastronomie Millimétrique, Saint Martin d'Hères, France
23 Department of Physics and Astronomy, University of Waterloo, Waterloo, Ontario, Canada
24 Waterloo Centre for Astrophysics, University of Waterloo, Waterloo, Ontario, Canada
25 Perimeter Institute for Theoretical Physics, Waterloo, Ontario, Canada
26 Institute of Space and Astronautical Science, Japan Aerospace Exploration Agency, Sagamihara, Japan

27 Department of Space and Astronautical Science, The Graduate University for Advanced Studies, SOKENDAI, Sagamihara, Japan
28 Dipartimento di Fisica e Astronomia, Università di Bologna, Bologna, Italy
29 Istituto di Radio Astronomia, INAF, Bologna, Italy
30 Department of Astronomy, Graduate School of Science, The University of Tokyo, Tokyo, Japan
31 Instituto Nacional de Astrofísica, Óptica y Electrónica, Puebla, Mexico
32 Academic Support Center, Kogakuin University of Technology and Engineering, Hachioji, Japan
33 Graduate School of Science and Technology, Niigata University, Niigata, Japan
34 Department of Space, Earth and Environment, Chalmers University of Technology, Onsala Space Observatory, Onsala, Sweden
35 Institute for Astrophysical Research, Boston University, Boston, MA, USA
36 Department of Electronics and Nanoengineering, Aalto University, Aalto, Finland
37 Metsähovi Radio Observatory, Aalto University, Kylmälä, Finland
38 Observatorio de Yebes, IGN, Yebes, Spain
39 National Radio Astronomy Observatory, Socorro, NM, USA
40 Department of Physics, Faculty of Science, Universiti Malaya, Kuala Lumpur, Malaysia
41 University of Vermont, Burlington, VT, USA.
42 East Asian Observatory, Hilo, HI, USA
43 University of Science and Technology, Daejeon, Republic of Korea
44 Institute of Astrophysics, Foundation for Research and Technology, Heraklion, Greece
45 Department of Physics, University of Crete, Heraklion, Greece
46 System Development Center, National Chung-Shan Institute of Science and Technology, Taoyuan, Taiwan, ROC
47 Institute of Astronomy and Astrophysics, Academia Sinica, Hilo, HI, USA
48 ASTRON, Dwingeloo, The Netherlands
49 Western University, London, Ontario, Canada
50 Graduate school of Science, Osaka Metropolitan University, Osaka, Japan
51 European Southern Observatory, Santiago, Chile
52 Leiden Observatory, University of Leiden, Leiden, The Netherlands

53 National Radio Astronomy Observatory, Charlottesville, VA, USA
54 Electronic Systems Research Division, National Chung-Shan Institute of Science and Technology, Taoyuan, Taiwan, ROC
55 Japan Aerospace Exploration Agency, Tsukuba, Japan
56 Department of Physics, National Sun Yat-Sen University, Kaohsiung City, Taiwan, ROC
57 Department of Physics, National Cheng Kung University, Tainan, Taiwan, ROC
58 Department of Physics, National Taiwan University, Taipei, Taiwan, ROC
59 Joint ALMA Observatory, Santiago, Chile
60 Green Bank Observatory, Green Bank, WV, USA
61 Astronomy Department, Universidad de Concepción, Concepción, Chile
62 Center for Astrophysics | Harvard & Smithsonian, Cambridge, MA, USA
63 Argelander-Institut für Astronomie, Universität Bonn, Bonn, Germany
64 Institut de Radioastronomie Millimétrique, Granada, Spain

**The nearby radio galaxy M87 is a prime target for studying black hole accretion and jet formation[1,2]. Event Horizon Telescope observations of M87 in 2017, at a wavelength of 1.3 mm, revealed a ring-like structure, which was interpreted as gravitationally lensed emission around a central black hole[3]. Here we report images of M87 obtained in 2018, at a wavelength of 3.5 mm, showing that the compact radio core is spatially resolved. High-resolution imaging shows a ring-like structure of $8.4^{+0.5}_{-1.1}$ Schwarzschild radii in diameter, approximately 50% larger than that seen at 1.3 mm. The outer edge at 3.5 mm is also larger than that at 1.3 mm. This larger and thicker ring indicates a substantial contribution from the accretion flow with absorption effects in addition to the gravitationally lensed ring-like emission. The images show that the edge–brightened jet connects to the accretion flow of the black hole. Close to the black hole, the emission profile of the jet-launching region is wider than the expected profile of a black -hole-driven jet, suggesting the possible presence of a wind associated with the accretion flow.**

On 14-15 April 2018, we performed very long baseline interferometry (VLBI) observations of M87 with the Global Millimetre VLBI Array (GMVA) complemented by the phased Atacama Large Millimetre/submillimetre Array (ALMA) and the Greenland Telescope (GLT) at a wavelength of 3.5 mm (86 GHz). The addition of the phased ALMA and GLT to the GMVA significantly improved the north-south resolution (by a factor of around 4) and baseline coverage in the direction perpendicular to the M87 jet. In Fig. 1, we show the resulting maps of M87, with a triple-ridged jet emerging from a spatially resolved radio core, which appears as a faint ring, with two regions of enhanced brightness in the northward and southward sections of the ring (Supplementary Information sections 2-4).

The most important feature of the image in Fig. 1a is the spatially-resolved radio core. With the nominal resolution of our VLBI array, we see two bright regions of emission oriented in the north-south direction at the base of the northern and southern jet rails (Fig. 1a). Motivated by an obvious minimum ("null") in the visibility amplitudes (Supplementary Figs. 10 and 11), we applied newly developed imaging methods that can achieve a higher angular resolution. This was done with and without subtracting the outer jet emission, to have a robust assessment of the parameters of the core structure (Supplementary Information section 3). From these images and by comparing ring- and non-ring-like model fits in the visibility domain, we conclude that the structure seen with the nominal resolution is the signature of an underlying ring-like structure with a diameter of $64^{+4}_{-8}$ μas (Supplementary Information), which is most apparent in slightly super-resolved images (Fig. 1b,c). Adopting a distance of $D$=16.8 Mpc and a black hole mass of $M$=6.5×10$^9$ $M_\odot$ (where $M_\odot$ is the solar mass)[4], this angular diameter translates to a diameter of $8.4^{+0.5}_{-1.1}$ Schwarzschild radii ($R_s$=2$GM/c^2$, where $G$ is the gravitational constant, $M$ the black hole mass, and $c$ the speed of light). On the basis of imaging analysis and detailed model fitting, we found that a thick ring (width $\gtrsim$20 μas) is preferred over a thin ring (Supplementary Information). We note that the observed azimuthal asymmetry in the intensity distribution along the ring-like structure may (at least partly) be due to the effects from the non-uniform ($u,v$) coverage (Supplementary Information section 4), which also would explain the north-south dominance of the emission in the ring. Moreover, this double structure may also mark the two footpoints of the northern and southern ridge of the edge-brightened jet emission, which is seen further downstream. We note that previous GMVA observations[5] – without the inclusion of ALMA and the GLT – had a lower angular resolution, which was insufficient to show the ring-jet connection, but it is seen in the present images. We further note that the published 1.3 mm images did not reveal the inner jet emission because of ($u,v$)-coverage limitations[6] (cf. refs.[6-8]).

The ring-like structure observed at 3.5 mm differs from the one seen at 1.3 mm. The ring diameter at 3.5 mm ($64^{+4}_{-8}$ μas) is about 50% larger than at 1.3 mm (42±3 μas, ref.[4]). This larger size at 3.5 mm is not caused by observational effects (for example, calibration or (*u,v*) coverage) and is already obvious from the (*u,v*)-distance plot of the visibilities (Supplementary Figs. 10 and 11). We note that the location of the visibility minimum, which scales inversely with the ring size, at 3.5 mm is at around 2.3 Gλ (Supplementary Information section 6). At 1.3 mm the first visibility minimum is seen at a significantly larger (*u,v*) distance of about 3.4 Gλ for the Event Horizon Telescope (EHT) data[9]. We find that the brightness temperature of the ring-like structure at 3.5 mm is approximately (1-2)×$10^{10}$ K and the total compact flux density is roughly 0.5-0.6 Jy (Supplementary Table 2).

The reported fine-scale structure of the M87 jet base is substantially different from the classic morphology of radio-loud active galactic nuclei, characterized by a compact, unresolved component (core), from which a bright, collimated jet of plasma emanates and propagates downstream. Figure 1 shows a spatially resolved radio core with a ring-like structure and a triple ridge jet structure[10] emerging to the west, with sharp gaps of emission between the ridges. Such a triple-ridge structure has been seen on larger scales ($\gtrsim$ 100 *R*s) in previous observations[5]. The location of the central ridge, which has an intensity about 60% of that of the outer jet ridges, suggests the presence of a central spine, which emerges from the ring centre. The jet expands parabolically along a position angle of approximately -67° (Supplementary Information), which is consistent with the jet morphology seen in previous studies[5]. Although previous images at 7 mm and 3.5 mm show some evidence for counter jet emission[5,11], we did not find any significant emission from a counter-jet in this 2018 observation (upper limit of ~ 1mJy/beam within 0.1-0.3 mas), possibly owing to its low brightness and limitations in the dynamical range.

Because we observed a ring-like structure, it is natural to assume that the black hole is located at its centre. Given the measured brightness temperature of about $10^{10}$ K being typical for active galactic nuclei cores, synchrotron emission is believed to be responsible for the 3.5 mm ring-like structure. At 1.3 mm, it has been shown that the emission is always strongly lensed into the observed ring shape, regardless of whether it originates near the equatorial plane associated with the accretion flow or the funnel wall jet (jet sheath)[12]. As shown below, our new observations at 3.5 mm can now constrain the spatial location and energy distribution of the electrons that are responsible for the millimetre emission.

The 2017 EHT observations have confirmed the nature of the accreting black hole in M87 to be in the low-Eddington regime, which is well described by a radiatively inefficient accretion flow (RIAF)[1,12]. On the basis of these studies, we model the spectral energy distribution and morphology of the horizon-scale structure assuming the emission is dominated either by the jet or by the accretion flow. This is done by applying a general relativistic radiative transfer to general relativistic magnetohydrodynamic simulations for an RIAF surrounding a rotating black hole (Supplementary Information section 9). The boundary between the accretion flow and jet is defined as the surface where the magnetic energy density equals the rest-mass energy density of the fluid (that is, $b^2/\rho c^2=1$; where $b$ is magnetic field strength, $\rho$ the plasma mass density and $c$ the speed of light). In the funnel region, $b^2/\rho c^2>1$ and synchrotron emission from electrons with a power-law energy distribution is assumed. Otherwise $b^2/\rho c^2<1$ and synchrotron emission from electrons with a Maxwellian energy distribution is considered.

The properties of the non-thermal synchrotron model (from the jet) and the thermal synchrotron model (from the accretion flow) are normalized to fit the core flux density at 1.3 mm observed by the EHT[12]. For both models, the plasma around the black hole is optically thin at 1.3 mm. The resultant model images (Fig. 2e,f) are consistent with the observed morphology in terms of flux density, ring diameter, and width (Fig. 2d). In both models, the ring-like structure observed at 1.3 mm is dominated by lensed emission around the black hole.

At 3.5 mm, the plasma in both models becomes optically thick because of synchrotron self-absorption, resulting in a ring-like structure (Figs. 2b,c), diameter of which is larger than that at 1.3 mm. However, owing to the different emissivity and absorption coefficients for thermal and non-thermal synchrotron emission[13], the diameter of the resulting ring-like structure at 3.5 mm for the non-thermal model (Fig. 2c) would be smaller ($\gtrsim$30%) than our observed value. By contrast, the thermal model (Fig. 2b) is able to produce a ring-like structure consistent with the 3.5-mm observations (Fig. 2a), suggesting that the thermal synchrotron emission from the accretion flow region plays an important part in the interpretation of the 3.5 mm GMVA observations.

We note a marginal variability of the 1.3 mm flux density between April 2017 and April 2018 (ref.[14]). With the assumption that the overall ring size (determined by the black hole) observed at 1.3 mm in April 2017 did not change significantly[3,15], a comparison of the 1.3-mm and 3.5-mm images with the model predictions allows us to conclude that the larger ring size at 3.5 mm indicates the detection of an accretion flow, which is affected by synchrotron self-absorption (opacity) effects.

Our 2018 images allow us to study the jet collimation below the roughly 0.8 mas (about 100 $R$s) scale in detail (Fig. 3). We note a change in the parabolic expansion near the ring ($\lesssim$ 0.2 mas, Region I), where the measured jet widths forms a plateau and becomes larger than the parabolic jet profile seen further downstream ($\gtrsim$ 0.2 mas; Regions II and III)[5,16,17].

The observed parabolic shape is consistent with a black-hole-driven jet using the Blandford-Znajek[18] process[19]. We note that the Blandford-Znajek jet model can produce a quasi-symmetric structure of limb-brightened jet emission if the black hole spin is moderately large (a $\gtrsim$ 0.5), whereas the disk-driven jet model can not[20]. Following previous studies[19], we examine the envelope of the Blandford-Znajek jet (light-grey-shaded area, Fig. 3). The observed jet width in the innermost region (Region I in Fig. 3), however, is larger than this expected Blandford-Znajek jet envelope. We point out that a wide opening angle Blandford-Znajek jet launched from a strongly magnetized accretion flow (the so-called magnetically arrested disk)[21] may have difficulty in explaining this excess jet width. Therefore, such width-profile flattening suggests an extra emission component outside the Blandford-Znajek jet.

In addition to the jet, high-mass loaded, gravitationally unbound and non-relativistic winds have been found in RIAF simulations[22,23]. They are driven by the combination of centrifugal force[24] and gas and magnetic pressure[23] and are considered as an essential component collimating the Blandford-Znajek jet into a parabolic shape[19,25]. Non-thermal electrons accelerated by physical processes such as magnetic reconnection and shocks presumably exist in the wind. The synchrotron radiation of these non-thermal electrons may be responsible for this extra emission component[24] outside the Blandford-Znajek jet.

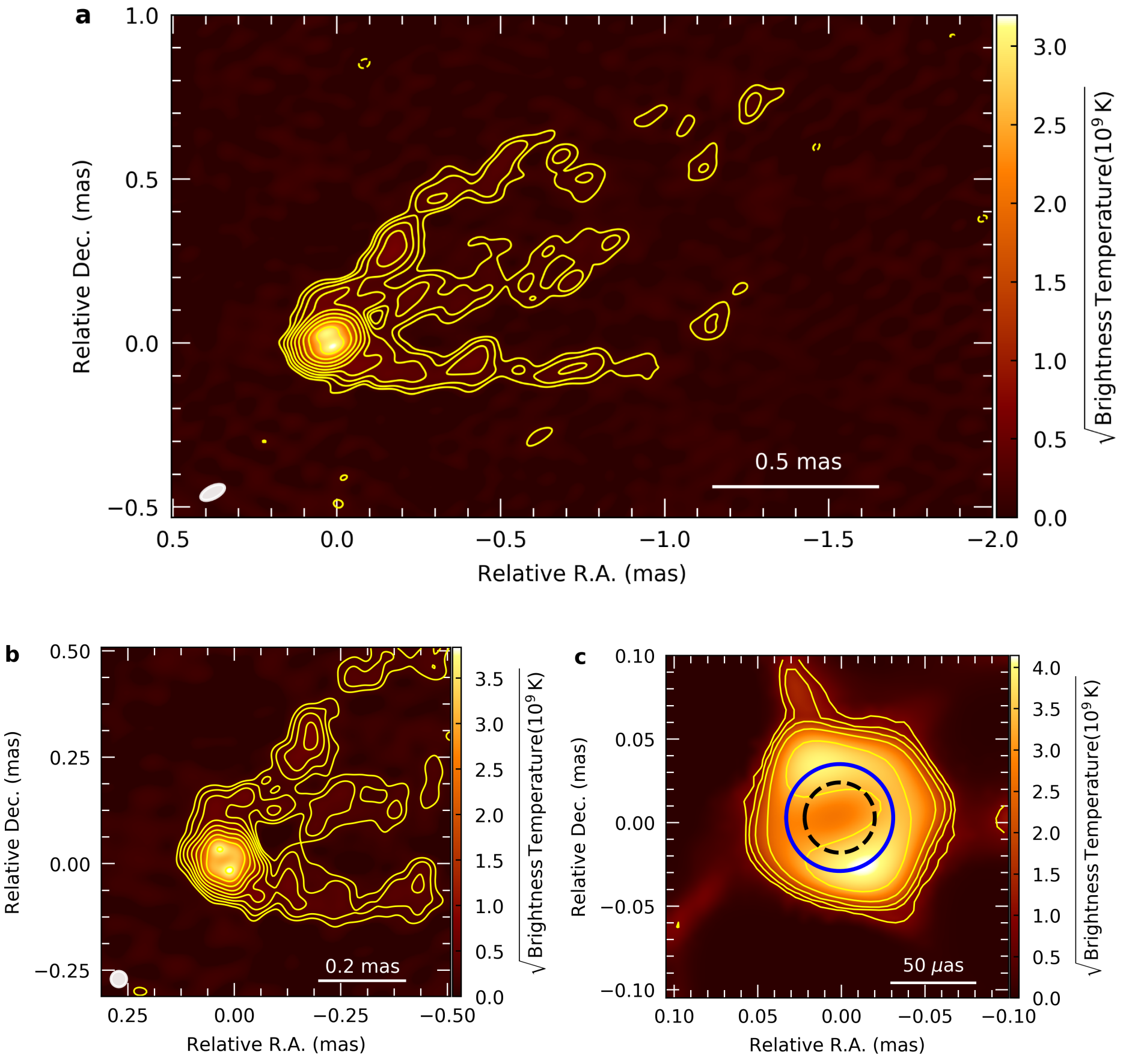

**Fig. 1|High-resolution images of M87 at 3.5 mm on 14-15 April 2018. a,** Uniformly weighted CLEAN image. The filled ellipse in the lower-left corner marks the restoring beam, which is an elliptical Gaussian fitted to the main lobe of the synthesised beam (FWHM: 79 μas × 37 μas, position angle p.a.= -63°). Contours show the source brightness in the standard radio convention of flux density per beam. The contour levels start at 0.5 mJy/beam and increase in steps of factors of 2. The peak flux density is 0.18 Jy beam$^{-1}$. **b,** The central region of the image as shown in **a**, but the image is now restored with a circular Gaussian beam of 37 μas size (FWHM), corresponding to the minor axis of the elliptical beam in **a**. The peak flux density is 0.12 Jy beam$^{-1}$. The contour levels start at 0.4 mJy/beam and increase in steps of factors of 2. **c,** A zoom into the central core region using regularised maximum likelihood (RML) imaging methods. Contours start at 4% of the peak and increase in steps of factors of 2. The solid blue circle of 64 μas in diameter denotes

the measured size of the ring-like structure at 3.5 mm, which is ~ 50% larger than the EHT 1.3 mm ring with a diameter of 42 μas (dashed black circle)[4]. For each panel, the colour map denotes the brightness temperature $T$ in kelvin, which is related to the flux density $S$ in jansky via the equation $T=\lambda^2(2k_B\Omega)^{-1}S$, where $\lambda$ is the wavelength, $k_B$ is the Boltzmann constant, and $\Omega$ is the solid angle, and is shown on a square-root scale. The CLEAN images are the mean of the best fitting images produced independently by team members, and the RML image is the mean of the optimal set of SMILI images (See Supplementary section 3).

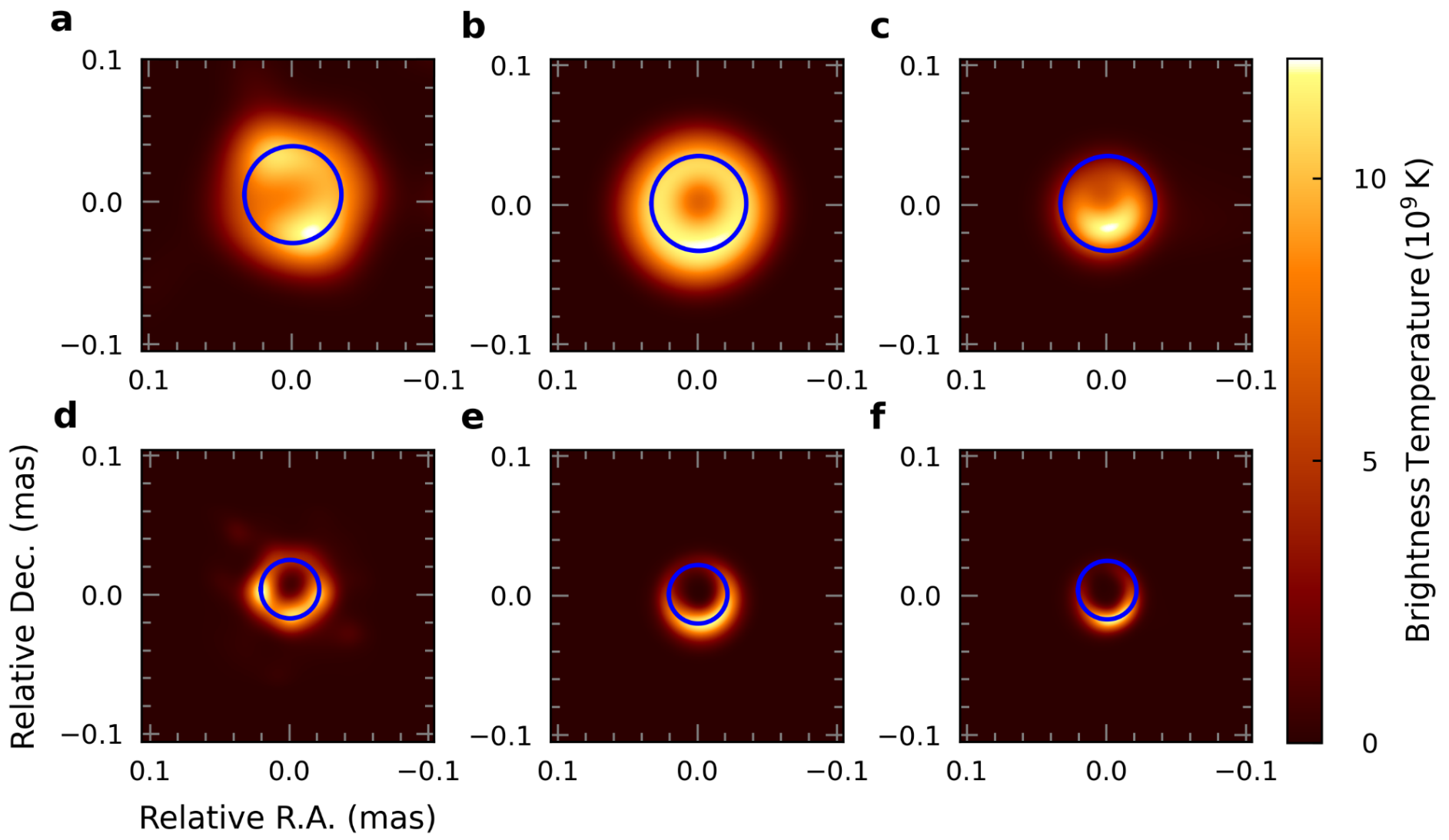

**Fig. 2| RML images (a, d) and model images (b, c, e, and f) at 3.5 mm (top) and 1.3 mm (bottom). a,** The 3.5 mm image obtained on 14-15 April 2018 is the same as in Fig. 1c but shown on a linear brightness scale**. b,e,** The thermal synchrotron model from the accretion flow assumes synchrotron emission from electrons with a Maxwellian energy distribution. **c,f,** The non-thermal synchrotron model from the jet region assumes synchrotron emission from electrons with a power-law energy distribution. **d,** The 1.3 mm EHT image obtained on 11 April 2017, reconstructed with the publicly-available data[9] and imaging pipeline[6] using the eht-imaging library[26]. Note that the differences in the azimuthal intensity distribution in the two observed images are probably because of time variability and/or blending effects with the underlying jet footpoints. Although the morphology of both models is consistent with the observations at 1.3 mm (**e** and **f**), the larger and thicker ring-like structure at 3.5 mm can be understood by the opacity effect at longer wavelengths[14], preferentially explained by thermal synchrotron absorption from the accretion flow

region (**b**). For comparison, reconstructed and simulated images are convolved with a circular Gaussian beam of 27 μas (3.5 mm) and 10 μas (1.3 mm) and are shown in a linear colour scale. The blue circle denotes the measured ring diameter of 64 μas at 3.5 mm and 42 μas at 1.3 mm.

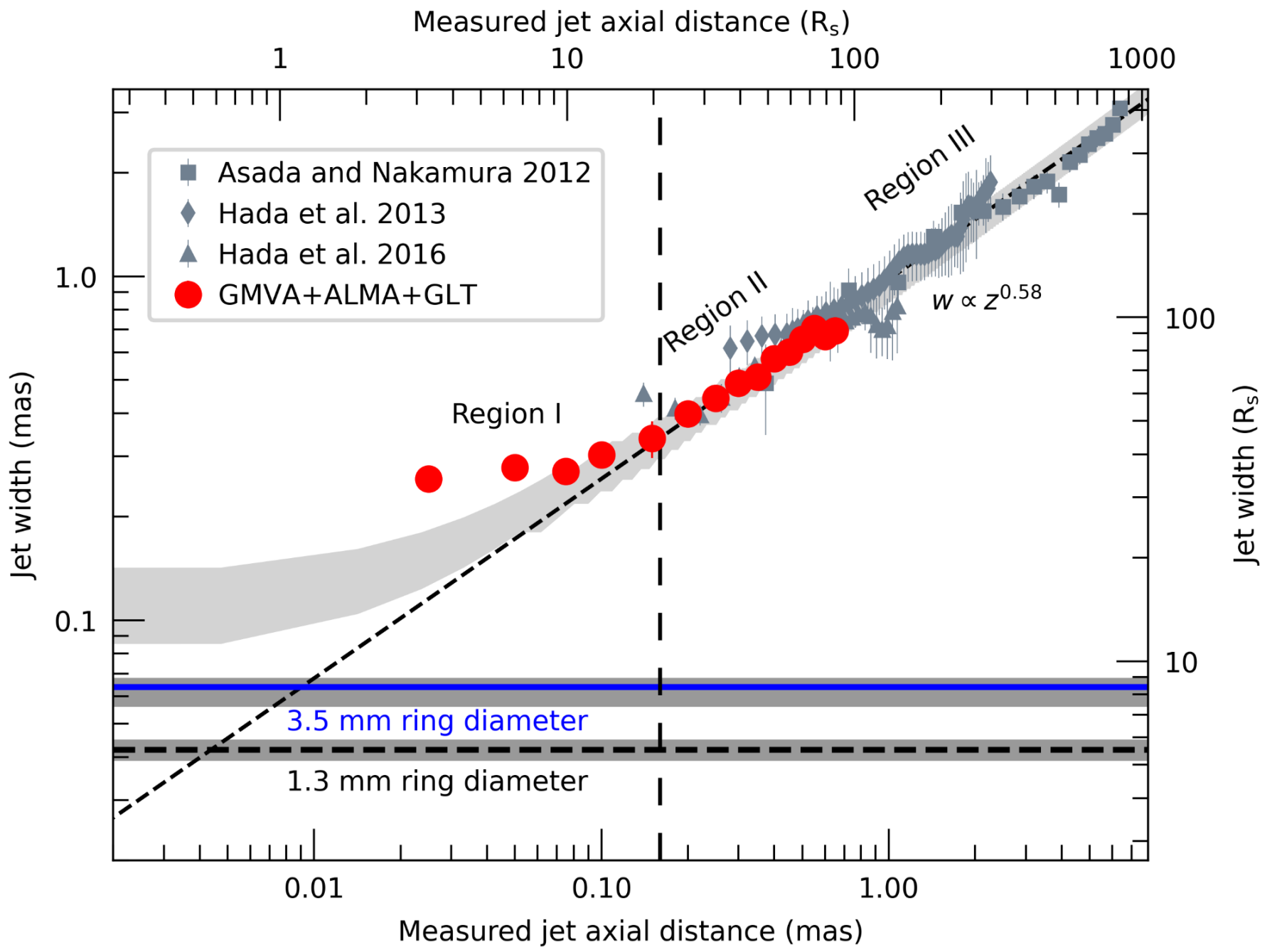

**Fig. 3| Jet collimation profile.** Red filled circles mark the measured jet transverse width for the observations reported here. The error bars (1 σ) are within the symbols (see Supplementary section 8 for more details on measuring the jet width). Grey filled squares, dots and triangles denote previous measurements of the width on larger scales[16,17,28], for which a power-law fit with a fixed power-law index of 0.58 is shown by the dashed line. The vertical dashed line marks the position at which the intrinsic half-opening angle $\theta$ of the fitted parabolic jet equals the jet viewing angle of $\theta_v = 17°$ (that is, boundary condition for a down-the-pipe jet[29]). The horizontal blue solid line marks the measured diameter of the ring at 3.5 mm, whereas the horizontal black dashed line marks the ring diameter measured with the EHT at 1.3 mm. In each case, the shaded area denotes the corresponding measurement uncertainty. The light-grey-shaded area denotes the outermost streamlines of the envelope of the parabolic jet from theoretical simulations (projected for $\theta_v = 17°$; ref.[30]) that are anchored at the event horizon[19] for a range of black hole spins (dimensionless spin parameters, $a = 0.0 – 0.9$). The lower and upper boundaries of this shaded area correspond to the highest ($a$=0.9) and lowest spin ($a$=0.0), respectively. As the jet footpoint is anchored at the event

horizon, some flattening of the jet width profile is expected near the black hole. This is further enhanced by geometrical projection effects in the region where the intrinsic jet half-opening angle ($\theta$) is larger than the jet viewing angle $\theta_v$. The quasi-cylindrical shape in Region I requires some change in the physical conditions to connect the innermost Blandford-Znajek jet from the event horizon to the upstream jet (Region II).

**Data availability**

The ALMA internal baseline data can be retrieved from the ALMA data portal under the project code 2017.1.00842.V. The calibrated VLBI data used in this paper are used in a continuing project but can be made available on reasonable request from the corresponding authors.

**Code availability**

Data processing and simulation softwares used in the paper including AIPS, DIFMAP, SMILI, and the eht-imaging library are publicly available. The perceptually uniform colormaps for image visualisation are available from the ehtplot library. The general relativistic magnetohydrodynamic simulation and general relativistic radiative transfer are performed with publicly available codes with HARM and ODYSSEY.

**Acknowledgements**

R.-S.L. is supported by the Key Program of the National Natural Science Foundation of China (grant No. 11933007), the Key Research Program of Frontier Sciences, CAS (grant No. ZDBS-LY-SLH011), the Shanghai Pilot Program for Basic Research – Chinese Academy of Sciences, Shanghai Branch (JCYJ-SHFY-2022-013), and the Max Planck Partner Group of the MPG and the CAS. R.-S.L thanks L. Blackburn, L. Chen, Y.-Z. Cui, L. Huang, R. S. de Souza, and Y. Mizuno for helpful discussions on data calibration and interpretation. J.P. acknowledges financial support through the EACOA Fellowship awarded by the East Asia Core Observatories Association, which consists of the Academia Sinica Institute of Astronomy and Astrophysics, the National Astronomical Observatory of Japan, Center for Astronomical Mega-Science, Chinese Academy of Sciences, and the Korea Astronomy and Space Science Institute. H.-Y.P. acknowledges the support of the Ministry of Education (MoE) Yushan Young Scholar Program, the Ministry of Science and Technology (MoST) under the Grant No. 110-2112-M-003-007-MY2, and the Physics Division, National Center for Theoretical Sciences. K.H. is supported by JSPS KAKENHI grant Nos. JP18H03721, JP19H01943, JP18KK0090, JP2101137, JP2104488, JP22H00157. J.-Y. Kim acknowledges support from the National Research Foundation (NRF) of Korea (grant no. 2022R1C1C1005255). I. M.-V. acknowledges support from Research Project PID2019-108995GB-C22 of Ministerio de Ciencia e Innovacion (Spain), from the GenT Project CIDEGENT/2018/021 of Generalitat Valenciana (Spain), and from the Project European Union NextGenerationEU (PRTR-C17I1). FY and HY are supported by Natural Science Foundation of China (grants 12133008, 12192220, 12192223) and China Manned Space Project (CMS-CSST-2021-B02). A.E.B. thanks the Delaney Family for their generous financial support via the Delaney Family John A. Wheeler Chair at Perimeter Institute. This work was supported in part by Perimeter Institute for Theoretical Physics. Research at Perimeter Institute is supported by the Government of Canada through the Department of Innovation, Science and Economic Development Canada and by the Province of Ontario through the Ministry of Economic Development, Job Creation and Trade. A.E.B. receives additional financial support from the Natural Sciences and Engineering Research Council of Canada through a Discovery Grant. S.K. acknowledges the program to support research activities of female researchers "Female Researchers Flowering Plan" from MEXT of Japan. The research at Boston University was supported in part by NASA Fermi Guest Investigator grant 80NSSC20K1567. H.N. is supported by JSPS KAKENHI grant No. JP18K03709 and JP21H01137. T.S. was partly supported by the Academy of Finland projects 274477, 284495, 312496, and 315721. P.dV. and B.T. thank the support from the European Research Council through Synergy Grant ERC-2013-SyG, G.A. 610256 (NANOCOSMOS) and from the Spanish Ministerio de Ciencia e Innovación (MICIU) through project PID2019-107115GB-C21. B.T. also thanks the Spanish MICIU for funding support from grants PID2019-106235GB-I00 and PID2019-105203GB-C21. This publication acknowledges project M2FINDERS that is funded from the European Research Council (ERC) under the European Union's Horizon 2020 research and innovation programme (grant agreement No 101018682). C.C. acknowledges support by the European Research Council (ERC) under the HORIZON ERC Grants 2021 programme under grant agreement No. 101040021. D.G.N. acknowledges funding from Conicyt through Fondecyt Postdoctorado (project code 3220195). This research has made use of data obtained with the Global Millimeter VLBI Array (GMVA), which consists of telescopes operated by the MPIfR, IRAM, Onsala, Metsähovi Radio

Observatory, Yebes, the Korean VLBI Network, the Greenland Telescope, the Green Bank Observatory (GBT) and the Very Long Baseline Array (VLBA). The VLBA and the GBT are facilities of the National Science Foundation operated under cooperative agreement by Associated Universities, Inc. The data were correlated at the VLBI correlator of the Max-Planck-Institut für Radioastronomie (MPIfR) in Bonn, Germany. This paper makes use of the following ALMA data: ADS/JAO.ALMA\#2017.1.00842.V. ALMA is a partnership of ESO (representing its member states), NSF (USA) and NINS (Japan), together with NRC (Canada), MOST and ASIAA (Taiwan), and KASI (Republic of Korea), in cooperation with the Republic of Chile. The Joint ALMA Observatory is operated by ESO, AUI/NRAO and NAOJ. The Greenland Telescope (GLT) is operated by the Academia Sinica Institute of Astronomy and Astrophysics (ASIAA) and the Smithsonian Astrophysical Observatory (SAO). The GLT is part of the ALMA-Taiwan project, and is supported in part by the Academia Sinica (AS) and the Ministry of Science and Technology (MOST) of Taiwan; 103-2119-M-001-010-MY2, 105-2112-M-001-025-MY3, 105-2119-M-001-042, 106-2112-M-001-011, 106-2119-M-001-013, 106-2119-M-001-027, 106-2923-M-001-005, 107-2119-M-001-017, 107-2119-M-001-020, 107-2119-M-001-041, 107-2119-M-110-005, 107-2923-M-001-009, 108-2112-M-001-048, 108-2112-M-001-051, 108-2923-M-001-002, 109-2112-M-001-025, 109-2124-M-001-005, 109-2923-M-001-001, 110-2112-M-003-007-MY2, 110-2112-M-001-033, 110-2124-M-001-007, 110-2923-M-001-001 and 110-2811-M-006-012. This research is partly based on observations with the 100-m telescope of the Max-Planck-Institut für Radioastronomie at Effelsberg. This work is partly based on observations carried out with the IRAM 30m telescope. IRAM is supported by INSU/CNRS (France), MPG (Germany) and IGN (Spain). This material is based upon work supported by the Green Bank Observatory which is a major facility funded by the National Science Foundation operated by Associated Universities, Inc. This work is partly based on observations carried out with the Yebes 40m telescope. The 40m radio telescope at Yebes Observatory is operated by the Spanish Geographic Institute (IGN, Ministerio de Transportes, Movilidad y Agenda Urbana). We acknowledge support from the Onsala Space Observatory national infrastructure for the provisioning of its facilities/observational support. Onsala Space Observatory receives funding through the Swedish Research Council via grant No 2017 – 00648. This publication makes use of data obtained at the Metsähovi Radio Observatory, operated by the Aalto University. This study makes use of VLBA data from the VLBA-BU Blazar Monitoring Program (BEAM-ME and VLBA-BU-BLAZAR; http://www.bu.edu/blazars/BEAM-ME.html), funded by NASA through the Fermi Guest Investigator Program.

**Author Contributions**

R.-S.L., K. Asada, T.P.K., and K.H initiated the project and have coordinated the research. P.H., M.I., M.-T.C., S.M., K.As., P.M.K., Y.-D.H., C.-C.H., D.K., P.A.R., T.J.N., N.A.P., and other engineers/technicians in East Asia have made the GLT available for our observations. K.As., T.P.K., R.-S.L, I.M.-V., J.P and E.R. have worked on the VLBI scheduling and coordination of the observations, analysis and calibration of the data. K. Akiyama, K. Asada., K.H., J.Y. Kim, T.P.K., R.-S.L., J.P., F.T., and A.L. have worked on image reconstruction and model fitting. K. Asada, J.L.G., K.H., J.Y.Kim, T.P.K., T.K., A.L. R.-S.L., M.N., J.P., H.-Y.P. and F.Y. have worked on the scientific interpretation of the results. All authors approved the paper and contributed to write the observing proposal, carry out the observations, produce and/or apply software tools for analysis and interpretation and contribute to the general interpretation of the data.

**Competing interests**

The authors declare no competing interests.

**Supplementary Information**

Supplementary Information is available for this paper.

**Author Information**

Correspondence and requests for materials should be addressed to Ru-Sen Lu (rslu@shao.ac.cn), Keiichi Asada (asada@asiaa.sinica.edu.tw), Thomas Krichbaum (tkrichbaum@mpifr-bonn.mpg.de), or Kazuhiro Hada (kazuhiro.hada@nao.ac.jp).

**Supplementary Information**

**A ring-like accretion structure in M87 connecting its black hole and jet**

Ru-Sen Lu[1,2,3,*], Keiichi Asada[4,*], Thomas P. Krichbaum[3,*], Jongho Park[4,5], Fumie Tazaki[6,7], Hung-Yi Pu[8,4,9], Masanori Nakamura[10,4], Andrei Lobanov[3], Kazuhiro Hada[7,11,*], Kazunori Akiyama[12,13,14], Jae-Young Kim[15,5,3], Ivan Marti-Vidal[16,17], José L. Gómez[18], Tomohisa Kawashima[19], Feng Yuan[1,20,21], Eduardo Ros[3], Walter Alef[3], Silke Britzen[3], Michael Bremer[22], Avery E. Broderick[23,24,25], Akihiro Doi[26,27], Gabriele Giovannini[28,29], Marcello Giroletti[29], Paul T. P. Ho[4], Mareki Honma[7,11,30], David H. Hughes[31], Makoto Inoue[4], Wu Jiang[1], Motoki Kino[14,32], Shoko Koyama[33,4], Michael Lindqvist[34], Jun Liu[3], Alan P. Marscher[35], Satoki Matsushita[4], Hiroshi Nagai[14,11,] Helge Rottmann[3], Tuomas Savolainen[36,37,3], Karl-Friedrich Schuster[22], Zhi-Qiang Shen[1,2], Pablo de Vicente[38], R. Craig Walker[39], Hai Yang[1,21], J. Anton Zensus[3], Juan Carlos Algaba[40], Alexander Allardi[41], Uwe Bach[3], Ryan Berthold[42], Dan Bintley[42], Do-Young Byun[5,43], Carolina Casadio[44,45], Shu-Hao Chang[4], Chih-Cheng Chang[46], Song-Chu Chang[46], Chung-Chen Chen[4], Ming-Tang Chen[47], Ryan Chilson[47], Tim C. Chuter[42], John Conway[34], Geoffrey B. Crew[13], Jessica T. Dempsey[42,48], Sven Dornbusch[3], Aaron Faber[49], Per Friberg[42], Javier González García[38], Miguel Gómez Garrido[38], Chih-Chiang Han[4], Kuo-Chang Han[46], Yutaka Hasegawa[50], Ruben Herrero-Illana[51], Yau-De Huang[4], Chih-Wei L. Huang[4], Violette Impellizzeri[52,53], Homin Jiang[4], Hao Jinchi[54], Taehyun Jung[5], Juha Kallunki[37], Petri Kirves[37], Kimihiro Kimura[55], Jun Yi Koay[4], Patrick M. Koch[4], Carsten Kramer[22], Alex Kraus[3], Derek Kubo[47], Cheng-Yu Kuo[56], Chao-Te Li[4], Lupin Chun-Che Lin[57], Ching-Tang Liu[4], Kuan-Yu Liu[4], Wen-Ping Lo[58,4], Li-Ming Lu[46], Nicholas MacDonald[3], Pierre Martin-Cocher[4], Hugo Messias[59,51], Zheng Meyer-Zhao[48,4], Anthony Minter[60], Dhanya G. Nair[61], Hiroaki Nishioka[4], Timothy J. Norton[62], George Nystrom[47], Hideo Ogawa[50], Peter Oshiro[47], Nimesh A Patel[62], Ue-Li Pen[4], Yurii Pidopryhora[3,63], Nicolas Pradel[4], Philippe A. Raffin[47], Ramprasad Rao[62], Ignacio Ruiz[64], Salvador Sanchez[64], Paul Shaw[4], William Snow[47], T. K. Sridharan[53,62], Ranjani Srinivasan[62,4], Belén Tercero[38], Pablo Torne[64], Efthalia Traianou[18,3], Jan Wagner[3], Craig Walther[42], Ta-Shun Wei[4], Jun Yang[34], Chen-Yu Yu[4]

1 Shanghai Astronomical Observatory, Chinese Academy of Sciences, Shanghai, People's Republic of China

2 Key Laboratory of Radio Astronomy, Chinese Academy of Sciences, Nanjing, People's Republic of China

3 Max-Planck-Institut für Radioastronomie, Bonn, Germany
4 Institute of Astronomy and Astrophysics, Academia Sinica, Taipei, Taiwan, ROC
5 Korea Astronomy and Space Science Institute, Daejeon, Republic of Korea
6 Simulation Technology Development Department, Tokyo Electron Technology Solutions, Oshu, Japan
7 Mizusawa VLBI Observatory, National Astronomical Observatory of Japan, Oshu, Japan
8 Department of Physics, National Taiwan Normal University, Taipei, Taiwan, ROC
9 Center of Astronomy and Gravitation, National Taiwan Normal University, Taipei, Taiwan, ROC
10 Department of General Science and Education, National Institute of Technology, Hachinohe College, Hachinohe City, Japan
11 Department of Astronomical Science, The Graduate University for Advanced Studies, SOKENDAI, Mitaka, Japan
12 Black Hole Initiative, Harvard University, Cambridge, MA, USA
13 Massachusetts Institute of Technology Haystack Observatory, Westford, MA, USA
14 National Astronomical Observatory of Japan, Mitaka, Japan
15 Department of Astronomy and Atmospheric Sciences, Kyungpook National University, Daegu, Republic of Korea
16 Departament d'Astronomia i Astrofísica, Universitat de València, Valencia, Spain
17 Observatori Astronòmic, Universitat de València, Valencia, Spain
18 Instituto de Astrofísica de Andalucía-CSIC, Granada, Spain
19 Institute for Cosmic Ray Research, The University of Tokyo, Chiba, Japan
20 Key Laboratory for Research in Galaxies and Cosmology, Chinese Academy of Sciences, Shanghai, People's Republic of China
21 School of Astronomy and Space Sciences, University of Chinese Academy of Sciences, Beijing, People's Republic of China
22 Institut de Radioastronomie Millimétrique, Saint Martin d'Hères, France
23 Department of Physics and Astronomy, University of Waterloo, Waterloo, Ontario, Canada
24 Waterloo Centre for Astrophysics, University of Waterloo, Waterloo, Ontario, Canada
25 Perimeter Institute for Theoretical Physics, Waterloo, Ontario, Canada

26 Institute of Space and Astronautical Science, Japan Aerospace Exploration Agency, Sagamihara, Japan
27 Department of Space and Astronautical Science, The Graduate University for Advanced Studies, SOKENDAI, Sagamihara, Japan
28 Dipartimento di Fisica e Astronomia, Università di Bologna, Bologna, Italy
29 Istituto di Radio Astronomia, INAF, Bologna, Italy
30 Department of Astronomy, Graduate School of Science, The University of Tokyo, Tokyo, Japan
31 Instituto Nacional de Astrofísica, Óptica y Electrónica, Puebla, Mexico
32 Academic Support Center, Kogakuin University of Technology and Engineering, Hachioji, Japan
33 Graduate School of Science and Technology, Niigata University, Niigata, Japan
34 Department of Space, Earth and Environment, Chalmers University of Technology, Onsala Space Observatory, Onsala, Sweden
35 Institute for Astrophysical Research, Boston University, Boston, MA, USA
36 Department of Electronics and Nanoengineering, Aalto University, Aalto, Finland
37 Metsähovi Radio Observatory, Aalto University, Kylmälä, Finland
38 Observatorio de Yebes, IGN, Yebes, Spain
39 National Radio Astronomy Observatory, Socorro, NM, USA
40 Department of Physics, Faculty of Science, Universiti Malaya, Kuala Lumpur, Malaysia
41 University of Vermont, Burlington, VT, USA.
42 East Asian Observatory, Hilo, HI, USA
43 University of Science and Technology, Daejeon, Republic of Korea
44 Institute of Astrophysics, Foundation for Research and Technology, Heraklion, Greece
45 Department of Physics, University of Crete, Heraklion, Greece
46 System Development Center, National Chung-Shan Institute of Science and Technology, Taoyuan, Taiwan, ROC
47 Institute of Astronomy and Astrophysics, Academia Sinica, Hilo, HI, USA
48 ASTRON, Dwingeloo, The Netherlands
49 Western University, London, Ontario, Canada
50 Graduate school of Science, Osaka Metropolitan University, Osaka, Japan

51 European Southern Observatory, Santiago, Chile
52 Leiden Observatory, University of Leiden, Leiden, The Netherlands
53 National Radio Astronomy Observatory, Charlottesville, VA, USA
54 Electronic Systems Research Division, National Chung-Shan Institute of Science and Technology, Taoyuan, Taiwan, ROC
55 Japan Aerospace Exploration Agency, Tsukuba, Japan
56 Department of Physics, National Sun Yat-Sen University, Kaohsiung City, Taiwan, ROC
57 Department of Physics, National Cheng Kung University, Tainan, Taiwan, ROC
58 Department of Physics, National Taiwan University, Taipei, Taiwan, ROC
59 Joint ALMA Observatory, Santiago, Chile
60 Green Bank Observatory, Green Bank, WV, USA
61 Astronomy Department, Universidad de Concepción, Concepción, Chile
62 Center for Astrophysics | Harvard & Smithsonian, Cambridge, MA, USA
63 Argelander-Institut für Astronomie, Universität Bonn, Bonn, Germany
64 Institut de Radioastronomie Millimétrique, Granada, Spain

**1. Observations**

M87 was observed by the Global Millimetre VLBI Array (GMVA) in concert with the phased Atacama Large Millimetre/submillimetre Array (ALMA) and the Greenland Telescope (GLT)[31] at 3.5 mm (86 GHz; GMVA project code ML005). During the observations, 32 12-metre ALMA dishes were phased up and the array was in the C43-3 configuration. The GMVA is composed of the eight Very Long Baseline Array (VLBA) antennas equipped with 3.5 mm receivers (i.e., without Saint Croix and Hancock), the Effelsberg 100 m telescope (EB), the Green Bank telescope (GB), the Metsähovi 14 m telescope (MH), the Onsala 20 m telescope (ON), the IRAM 30 m telescope (PV), and the Yebes 40 m telescope (YS). The data were recorded with a total bandwidth of 256 MHz per polarisation divided in 8 intermediate frequency bands (IFs), each of 64 spectral channels. YS observed in only one polarisation (left circular polarisation, LCP). The ~ 13 h track (~ 6 h with European telescopes, ~ 7 h with ALMA and the USA telescopes, and ~ 12 h with the GLT) included two calibration sources (3C273 and 3C279), which were observed every 20-40 minutes. The data were correlated with the VLBI correlator at the Max Planck Institute for Radio Astronomy using DiFX[32].

## 2. Post-correlation data reduction

**Polarisation conversion for ALMA and calibration of the GLT instrumental polarisation** In VLBI, most stations record their signals in circular polarisation, denoted as "RCP" (right-hand circular polarisation) and "LCP" (left-hand). The use of this polarisation basis has important implications for the calibration of the VLBI observables[33]. Even though the VLBI antennas have to detect the sky signal in a circular polarisation basis, the signal detection at the backend has to be done in linear polarisation (i.e., the basis used by the detector dipoles in the waveguides). The translation between these two bases of polarisation is done by inserting hardware (between the horn at the frontend and the receiver) that rotates the relative phase between two orthogonal linear polarisations by 90º. Such a phase rotation can be applied following different instrumental configurations. One option is to add a quarter wave plate (QWP) into the signal path. Another option is to have a waveguide-based phase rotator, which is the option used at the GLT. In either case, the two orthogonal linear polarisations that propagate after the phase rotator (and are later converted to electronic signals) carry the amplitudes and phases corresponding to the on-sky circular polarisations (i.e., the on-sky RCP and LCP waves are converted into X and Y by the phase rotator).

During our GMVA observations in 2018, two issues related to instrumental polarisation arose which required pre-processing of the data prior to the standard calibration. First, phased ALMA recorded the observed signals using a linear polarisation basis, while all the other stations recorded the signal using a circular polarisation basis. Consequently, the visibilities between ALMA and the other stations were correlated with a mixed polarisation mode[33]. These visibilities were converted into a pure circular basis based on the internal calibration of the ALMA interferometric data using the PolConvert algorithm[33,34]. Second, our observations were performed during the commissioning phase of the GLT, during which its waveguide-based phase-rotator (used to convert the circular-polarisation on-sky signals into linear polarisation for detection at the backend) was erroneously configured resulting in the application of a rotation of 45° (instead of 90º) between the polarisation channels. Such a rotation, which is equivalent to a polarisation leakage with a maximum amplitude between RCP and LCP, cannot be corrected using standard algorithms. Due to this rotation, the polarisation basis registered at the GLT was highly elliptical, rather than circular. In addition to this, the electronic gains of the GLT (which depend on the polarisation channel) add relative amplitudes and phases between the (already corrupted) RCP and LCP signals, thus complicating the calibration.

We removed the instrumental polarisation effects from the GLT by applying, in the frame-work of the Radio Interferometer Measurement Equation (RIME, ref.[35]), the theoretical polarisation-leakage matrix that would correct for the wrong phase applied with the phase rotator. This matrix, however, can only be applied after the application of the polarisation-wise gain matrices, which are not known a-priori and have to be estimated from the data. We determined these gains by separating them into a time variable gain matrix that is independent of the polarisation channel (so it commutes with the instrumental polarisation matrix and does not affect the GLT polarimetry) and another matrix that is stable in time and encodes the cross-polarisation gains. Once these gains were estimated via least-squares fitting, we computed the total (i.e., gains + polarisation) calibration matrix and applied it to all the GLT-related data. This approach is an adaptation of the "single-station" mode of PolConvert (see Eq. 16 in ref.[33]), but with a conversion equation different from the standard one (i.e., Eq. 18 in ref.[33]). In our case, the conversion is done with the following equation (following the same nomenclature as ref.[33]):

$$V_{\odot\odot}^{\mathrm{corr}} = C_{\odot+} H_{+\odot} \begin{bmatrix} G_{r/l} & 0 \\ 0 & 1 \end{bmatrix} V_{\odot\odot}^{\mathrm{orig}}$$

where “corr” refers to the corrected visibility matrix, “orig” to the matrix obtained from the DiFX correlation, $G_{r/l}$ is the cross-polarisation GLT gain, $H$ is the matrix that removes the phase rotation and converts to linear polarisation basis, and $C$ is the matrix that converts linear to circular (i.e., the product of $C$ and $H$ is a Jones matrix that corrects the wrong phase rotation applied by the phase rotator).

In Figure S1, we show the result of our polarimetric GLT correction for a representative fringe. As expected, the bulk of the signal is transferred from the mixed (RL, LR) to the corrected parallel hand (RR, LL) correlations (i.e., the ones that are proportional to the total intensity). Several consistency checks were done to assess the GLT polarimetry correction. On the one hand, we checked for the stability of the GLT cross-polarisation gains, which remained constant (within a few percent of slow variability) throughout the experiment. On the other hand, the amplitudes of RL and LR were checked to be lower than those of RR and LL, and the RR/LL phases were compared to the difference in parallactic angles between the GLT and the other antennas (the changes in the former should be closely related to the latter). The data passed all these consistency tests successfully.

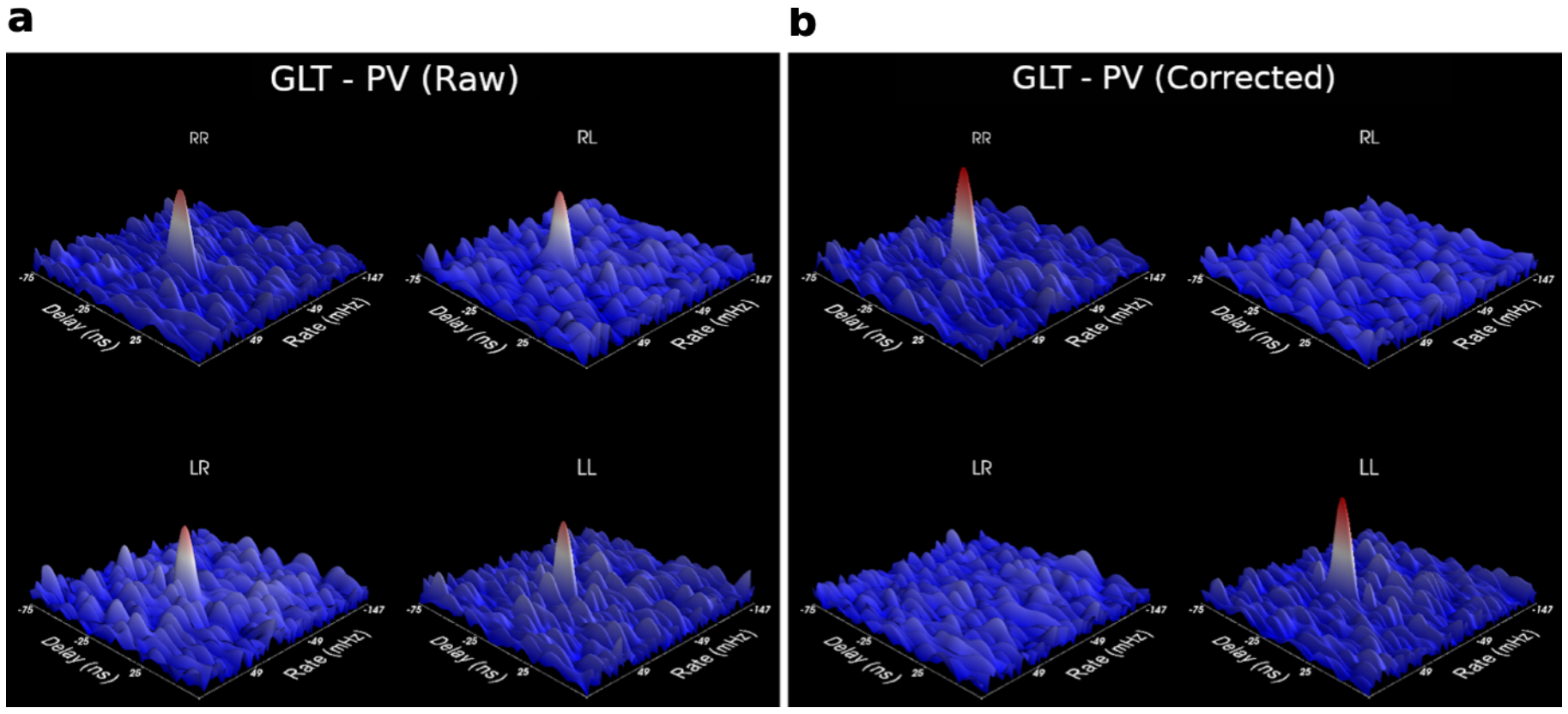

**Figure S1|Example of fringe amplitudes in a VLBI scan between GLT and PV, observing M87 for all 4 polarisation combinations (RR, RL, LR, LL)**. Only one subband centred at 86.3 GHz is shown (similar results are obtained for all the other subbands). **a**, after the DiFX correlation. **b**, after applying the GLT instrumental polarisation correction.

**Amplitude and phase calibration of the GMVA+ALMA+GLT data** Once the polarimetry issues were fixed, the data reduction was carried out using the NRAO Astronomical Imaging Processing System (AIPS, ref.[36]) in the standard manner. The initial fringe fitting of the data was done with the task FRING in a two-step procedure. In the first step, we used high SNR scans on the brighter calibrators to generate "manual" phase-cal information to remove relative phase offsets between IFs and a constant single-band delay relative to the reference antenna. In the second step, global fringe-fitting[37] was done to determine and remove residual single- and multi-band delays and fringe rates. In doing so, we assumed a point source model because the detailed source structure at this stage of the data analysis was still unknown. We then derived and applied bandpass corrections using calibrator scans and performed the a priori amplitude calibration in the standard manner, using measured system temperatures and gain curves. Atmospheric opacity corrections were applied based on station weather data and the fitting of the system temperature versus air mass. We then derived an initial source structure model for M87, which then was used in a second round of global fringe-fitting on M87 to lower the detection SNR (with a SNR cutoff of 4.3) and improve the fringe detection quality by narrowing the fringe search windows in rate and delay. After this, the data were then averaged over frequency before imaging. We detected fringes to M87 with all participating stations except for MK (VLBA) due to sensitivity limitations and non-optimal weather conditions in Hawaii (fringes to MK were detected, however, for the brighter calibrators). In Figure S2, we show the $(u,v)$-coverage for the data with fringe detections to M87. We also show examples of the measured closure phases as a function of time for the ALMA-GB-PV and ALMA-GB-GLT triangles. We note a minimum in the correlated flux density at ~ 2.3 Gλ (see also Figures S10 and S11) , a clear phase jump from ~ 0° to ~ 180° across the identified minimum, and closure phases of ~ 180° on ALMA-USA-Europe and ALMA-USA-GLT triangles.

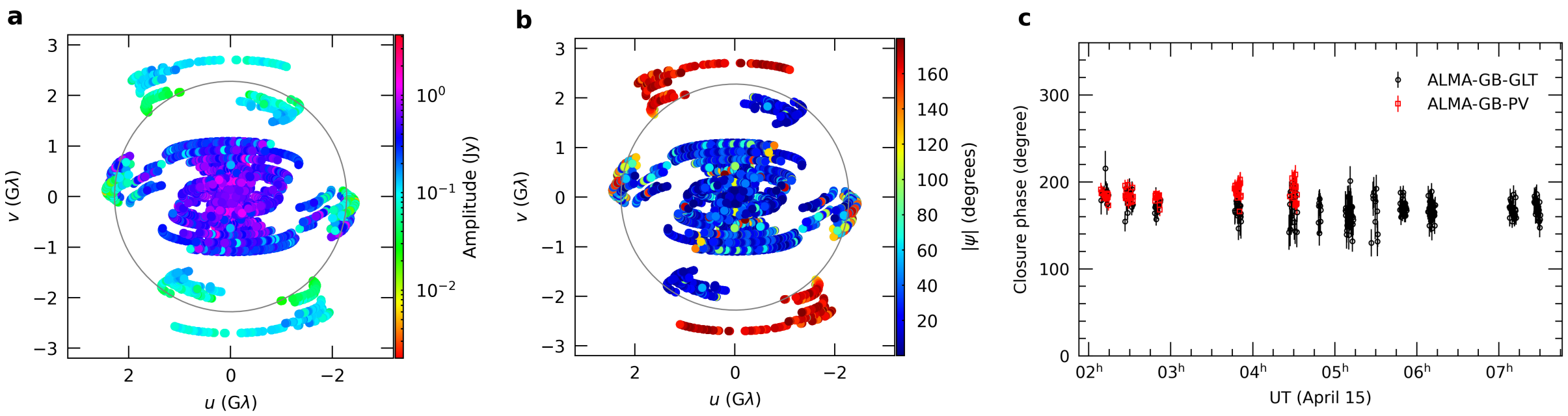

**Figure S2| $(u,v)$-coverage and closure phase of M87. a,** $(u,v)$-coverage plot colour-coded by self-calibrated visibility amplitudes and **b,** by the absolute value of phases |φ|. For each *uv*-plot, a circle with a radius of 2.3 Gλ, at which the visibility amplitude reaches minimum (visibility null), is drawn. This visibility null corresponds to a diameter of ~ 69 μas for an infinitesimally thin ring in the image plane. **c,** exemplary closure phases measured on two wide-open triangles (ALMA-GB-GLT and ALMA-GB-PV). Error bars represent $1\sigma$.

## 3. Imaging

Reconstruction of an image from interferometric data is a well-known ill-posed inverse problem. Interferometric imaging algorithms can be broadly categorised into the inverse methods (e.g., the widely-used, traditional CLEAN deconvolution algorithm[38]) or forward modelling methods (e.g., regularised maximum likelihood, RML, methods). In RML imaging, the basic concept is to find the most probable image that minimises a weighted sum of the chi-squared values ($\chi^2$) and (prior) regularisation functions. We refer the reader to ref.[6] and references therein for more details.

**Gain correction** At millimetre wavelengths, accurate amplitude calibration becomes more difficult due to temporal changes in sky opacity, less well-known antenna gains and gain-elevation effects, as well as antenna mechanical effects that are difficult to constrain (e.g., residual pointing and focus offsets). As a result, scans with low amplitudes are more often seen than in cm-VLBI, in particular, for baselines involving stations that were built for observations at longer (centimetre) wavelengths.

To correct for these calibration errors, we derived time-dependent station gain correction factors using the self-calibration algorithm which is implemented in the Difmap software[39], making use of CLEAN[38]. For doing this, the calibrated data were first coherently averaged in 10-second bins and outliers were flagged. Imaging (deconvolution) and phase and amplitude self-calibration were then done in an iterative manner. We performed several iterations of phase only self-calibration, followed by amplitude self-calibration iterations. For the amplitude self-calibration, we incrementally decreased the solution intervals from the entire observation duration down to the scan length (typically 6 minutes). This was done independently by 6 authors/teams.

We found that the residual errors associated with this iterative amplitude calibration do not significantly affect the dominant source features. Consequently, the amplitude self-calibration solutions show good overall consistency among all images created by different team members. The overall correction trends are also consistent with those for the two calibrators, which were observed repeatedly in short scans during the whole experiment (see Figure S4 in next section). Motivated by this consistency, we chose to correct the amplitudes of the calibrated data for the VLBA and the GBT based on the average of the gain corrections derived from the individual images. In doing so, we have assumed that the scans with the lowest gain correction factors for each station were not affected much by pointing errors and thus represent the true visibility amplitudes of our target. Implicitly, we scaled the gain corrections up such that the two lowest factors that correspond to the two best scans for each station became unity on average. After this correction, we found that the amplitudes on short baselines are in good agreement with near in time VLBI observations of the Korean VLBI Network (KVN) at 3.5 mm[40], which justifies the need for the scaling of the antenna gains. We then use these pre-corrected data for further imaging.

**Imaging with the CLEAN method** We performed a second-round of hybrid imaging and self-calibration in Difmap, now with the pre-corrected data. The individual images obtained by the different imaging team members are in good agreement with each other. The reduced chi-squares of the images for closure phases and log closure amplitudes are all below 1.3 and 1.4, respectively, indicating that the images fit the data and represent the source structure very well. The final and representative CLEAN image of M87 was then obtained by averaging the best fitting images (Fig. 1a,b in the main text).

**Imaging with the Regularised Maximum Likelihood (RML) method** As demonstrated in recent work[41,42], RML methods provide an alternative way of imaging, which can be less affected by the systematics inherent to deconvolution of incomplete ($u,v$)-coverage. Through the use of RML imaging it is possible to obtain high-fidelity images with better resolution than the nominal diffraction limit[43]. Here, we adopted the SMILI package[44,45] for RML imaging.

Since the CLEAN method recovered the extended jet structure on 0.1 - 1.0 mas scales relatively well, our primary motivation for the RML imaging was to confirm and image the spatially resolved core structure in more detail and at a higher resolution. In the first approach of SMILI imaging, we subtracted the extended jet structure from the CLEAN self-calibrated visibility data to focus on the imaging of the VLBI core of M87. In the second approach, we performed SMILI imaging of the amplitude pre-corrected data directly. In Figure S3 and Fig. 1c, we show the mean of the optimal set of images from each approach.

**(1) Imaging with jet-subtracted data** We first calculated the extended jet model visibilities using the CLEAN components located at distances from the centre of the core larger than $r_{\rm cut}$=50, 75, and 100 μas. The centre position was defined to be the midpoint of a two-point-source model fitted to the self-calibrated visibilities using a non-linear least squares method (task MODELFIT in Difmap[39]). We then subtracted the extended jet model visibilities from the data for each CLEAN self-calibration solution and for each cutoff distance. Thus, we produced 15 versions of the jet-subtracted data and used them as input for the subsequent SMILI imaging.

We first performed SMILI imaging on a small grid of parameters to check the effect of distance cutoff ($r_{\rm cut}$). We found that the images obtained from the datasets with $r_{\rm cut}$=75 and 100 μas appeared similar, while images from the dataset with $r_{\mathrm{cut}}$=50 showed more differences and appeared less consistent, indicating the distance cutoff was too small. In the end, we chose $r_{\mathrm{cut}}$=75 μas as cutoff for the final imaging on a larger grid of parameters.

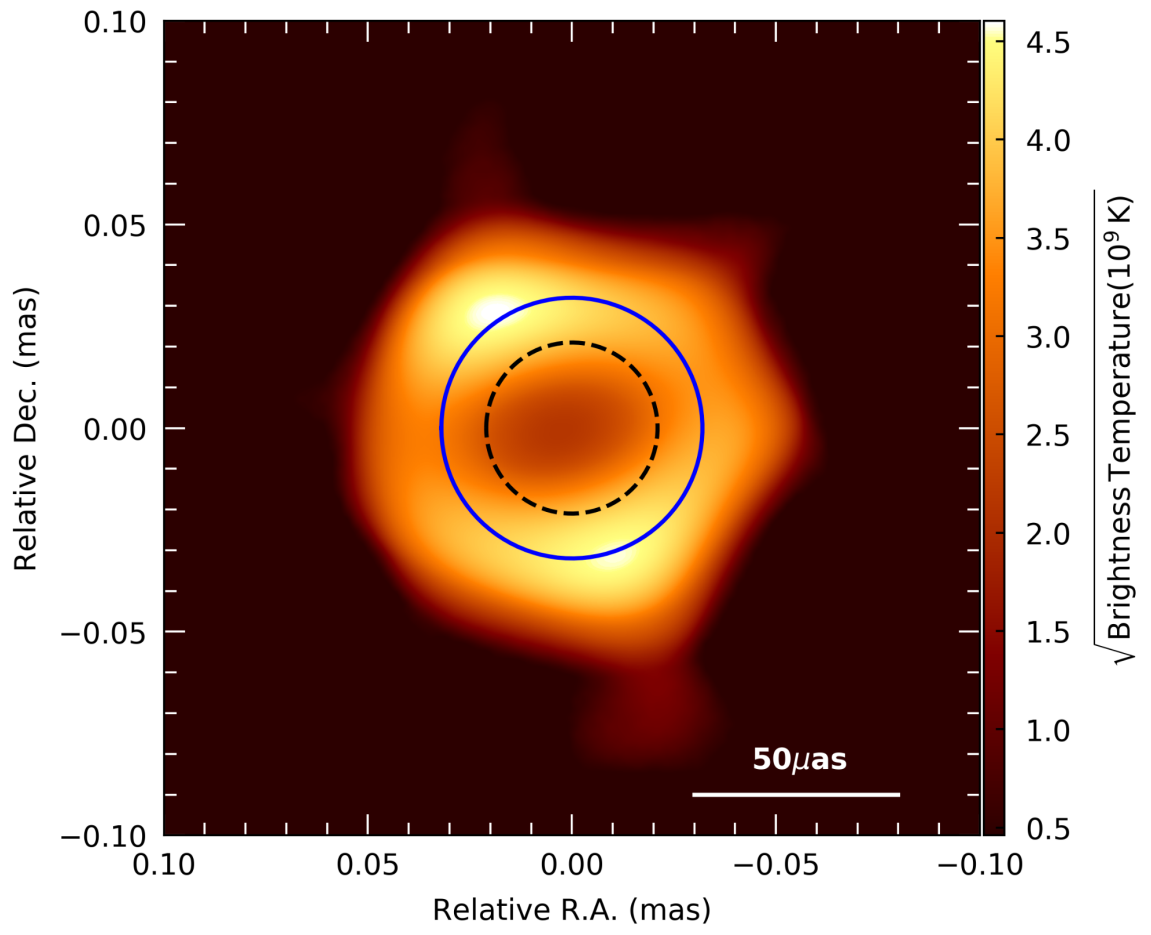

**Figure S3| Average image of the optimal set of images for the jet subtracted data.** The solid blue circle has a diameter of 64 μas and denotes the measured size of the ring-like structure at 3.5 mm. The dashed black circle denotes the ring diameter of 42 μas measured by the EHT at 1.3 mm one year earlier, in April 2017.

For each image reconstruction, we used pseudo Stokes I visibilities derived from individually self-calibrated data sets. Pseudo Stokes I visibilities are the weighted mean of parallel-hand visibilities of LL and RR polarisations and it avoids flagging visibilities on baselines that sample only one of the two polarisations. The field of view was set to 100 × 100 pixels with a pixel size of 2 μas. The total flux density was fixed based on the CLEAN imaging after subtracting the jet components. We utilised weighted-L1 (wL1, which favours sparsity in the image domain), total squared variation (TSV, which favours smooth edges), and total variation (TV, which favours sparsity in the image gradient domain) regularizers for sparse imaging (see ref.[6] for their mathematical definitions). A circular Gaussian image was introduced for weighted-L1 to increasingly penalise pixel intensities farther from the image centre. The imaging parameter grid consists of Gaussian FWHMs of 70, 80, 100, and 120 μas and regularisation parameters of wL1, TSV, and TV of [10,1,0.1], [100,10,1], and [100,10,1], respectively. In total, we explored a set of 540 parameters to reconstruct the core structure. In Figure S3, we show the mean of the images from the optimal parameter sets (245 images in total), for which each image, when combined with the subtracted jet model, shows better chi-square values in closure phase and log closure amplitude than the original CLEAN image.

**(2) Imaging the full data** We also performed SMILI imaging of the amplitude pre-corrected data using visibility amplitudes, closure phases, and log closure amplitudes. Systematic errors are added to the visibility amplitude to represent the gain errors that remained after the pre-correction: (i) 20% for ALMA, PV, and GLT, whose gains were relatively well determined, and (ii) 20-40% for the other stations, depending on their calibration uncertainty. To complete the gain corrections and obtain the final images, imaging and self-calibration were iterated three times. While the scan averaged data was used for imaging, the original 10-second integrated data was used for self-calibration. To account for possible discrepancies in the RR and LL gains, at the beginning of the iteration, we reconstructed an image from the RR and LL data, and also solved for the R and L gains independently. In the second and the last iterations, we used pseudo Stokes-I for the image reconstruction, since possible R-L gain offsets were already removed in the first self-calibration. In the image reconstruction, the total flux density was fixed to 0.6, 0.7, or 0.8 Jy. We used regularizers of wL1, TV, TSV, and relative entropy (maximum entropy method; MEM).

The hyper parameter of the wL1 term was fixed at 1.0. For the other hyper parameters, we tried multiple values with TSV, TV, and relative entropy of [-1, 10, 100, 1000], [-1, 10, 100, 1000], and [-1, 0.00001, 0.0001, 0.001], respectively. The wL1 regularisation requires an image prior, and we used a preliminary RML image with Gaussian convolution of 50, 100, and 200 μas FWHM. After three iterations of imaging and self-calibration, the final image set was obtained. Since the chi-squared values of all images now were reasonably low, a chi-squared cutoff was not applied. We then selected the optimal set of images under the condition that the gain offsets of ALMA, GLT, and PV are less than 15%. The resulting optimal set of images (216 of the 1728 generated images) have a total flux density in the range of 0.7 - 0.8 Jy. We adopted the average of the optimal set of images as a representative image (Fig. 1c).

## 4. Image validation

**Self-calibration gains and validation with the calibrator 3C273** We checked the accuracy and consistency of the amplitude calibration with self-calibration imaging of 3C273, which was included in this experiment as one of the two calibrators (the other being 3C279). We mainly used this source because the GLT did not observe 3C279 due to its low declination. We present the gain correction factors, the inverse of the derived antenna gains, of M87 and 3C 273 for each scan in Figure S4. We took the median of the gain correction factors from self-calibration with the best fitting CLEAN images of M87. We note the median gain correction factor for each station as $|1/G_{med}|$. We calculated the standard deviation of the gain correction factors between neighbouring (within 10 minutes) scans of the two sources and obtained the median value of the standard deviations for each station, which is noted as $\sigma_{M87\text{-}3C273}$. The gain correction factors from the two sources show very consistent trends with each other with $\sigma_{M87\text{-}3C273}$ typically less than about 20% for stations having a good number of detections on both sources for neighbouring scans. Based on this analysis, we conclude that the amplitude calibration is accurate to within ~ 20-30%.

The participation of ALMA and the GLT significantly improved the baseline coverage to 3C273 (Figure S5a). Our image at 3.5 mm shows an extended jet towards the south-west (Figure S5b), consistent with the jet orientation seen with a quasi-simultaneous 7 mm VLBA image from the Boston University VLBA monitoring program (Figure S5c). The peak brightness temperature of the jet at 3.5 mm is $\gtrsim 10^{11}$ K.

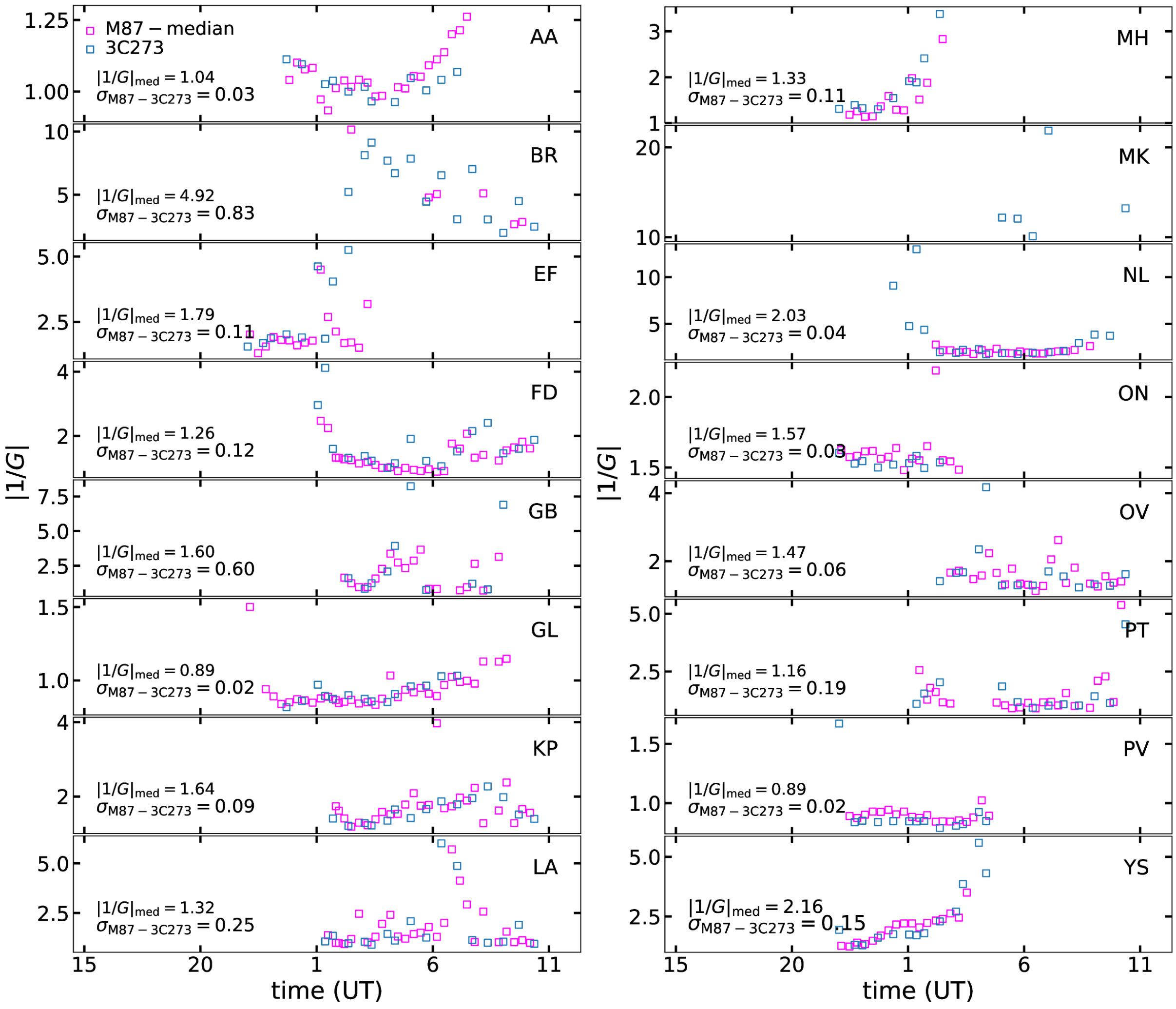

**Figure S4 | Comparison of station gain correction factors.** These factors are derived from the CLEAN models of M87 (magenta) and the calibrator 3C 273 (blue).

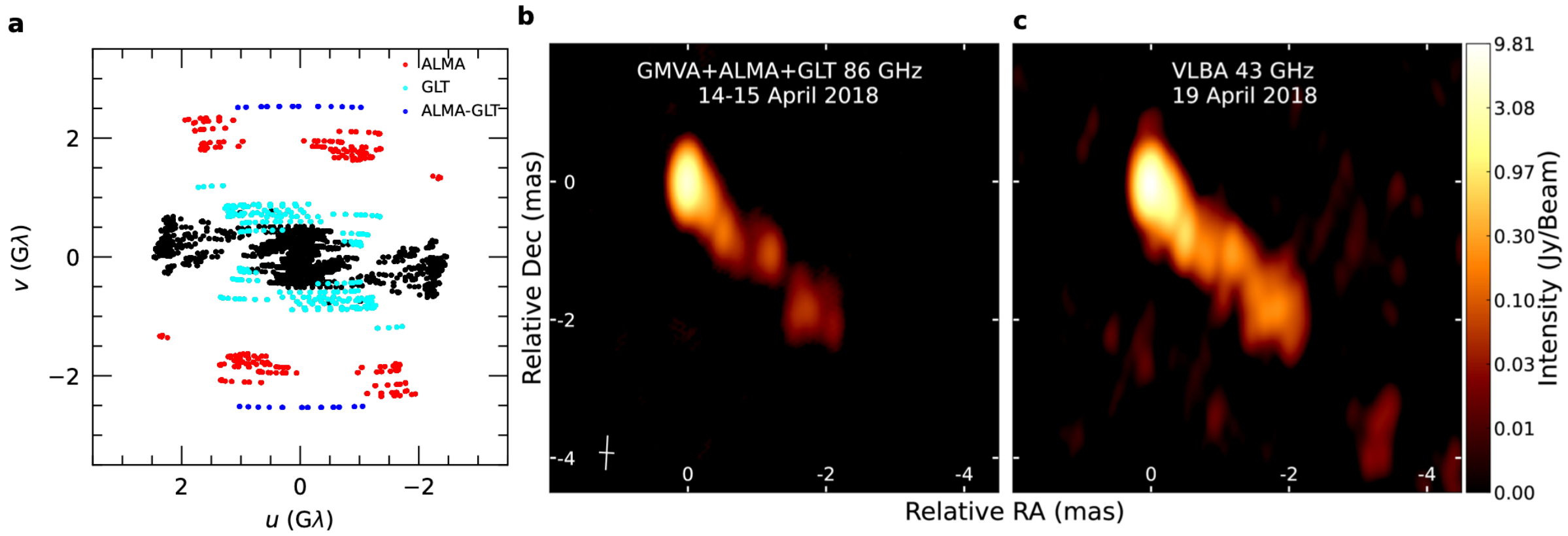

**Figure S5 | (*u*,*v*)-coverage and CLEAN images of the calibrator 3C273. a,** The participation of phased ALMA and GLT greatly improved the (*u*,*v*)-coverage to 3C273 (highlighted). **b,** Total intensity image of 3C 273 obtained from our GMVA+ALMA+GLT observations at 3.5 mm on 14-15 April 2018. **c**, VLBA image of 3C 273 at 7 mm obtained from observations on 19 April 2018. Both images are restored with the same synthesised beam of FWHM of 0.46 x 0.20 mas, for the major and minor axis respectively. The beam orientation at a position angle of -3.1° is indicated by the white cross on the bottom left of **b**.

**Dependence of the ring-like structure on SMILI regularizers** During the imaging, we found that all SMILI and super-resolution CLEAN images show a central brightness depression indicating a ring-like structure. Furthermore, our visibility domain model-fitting analysis, which is independent of resolution and beam convolution effects, confirms the presence of this ring-like structure with a finite width (Supplementary Information section 6). Such a morphology is supported by the prominent minimum (visibility null) with a clear phase jump from ~ 0° to ~ 180°. The increase in the visibility amplitudes between ~ 2.3 Gλ and ~ 3 Gλ (Figures S10 and S11) also favours a ring-like structure over other models without a central depression, e.g., a simple or double Gaussian or a uniform disk.

On the other hand, the azimuthal details of the ring-like structure are likely affected by the (*u*,*v*)-coverage (see the 'Synthetic data imaging tests' section below) and should therefore be interpreted with caution. Figure S6 shows the dependence of the ring-like structure on the chosen weighting for particular SMILI regularizers. We note that depending on the weighting of the edge-smoothing regularizer (total squared variation, TSV), the central C-shaped emission region appears more blurred and the central flux depression becomes less prominent. However, for smaller TSV values, a ring-like structure becomes most prominent. With these caveats in mind, we conclude that our observations provide strong evidence for a ring-like structure at 3.5 mm.

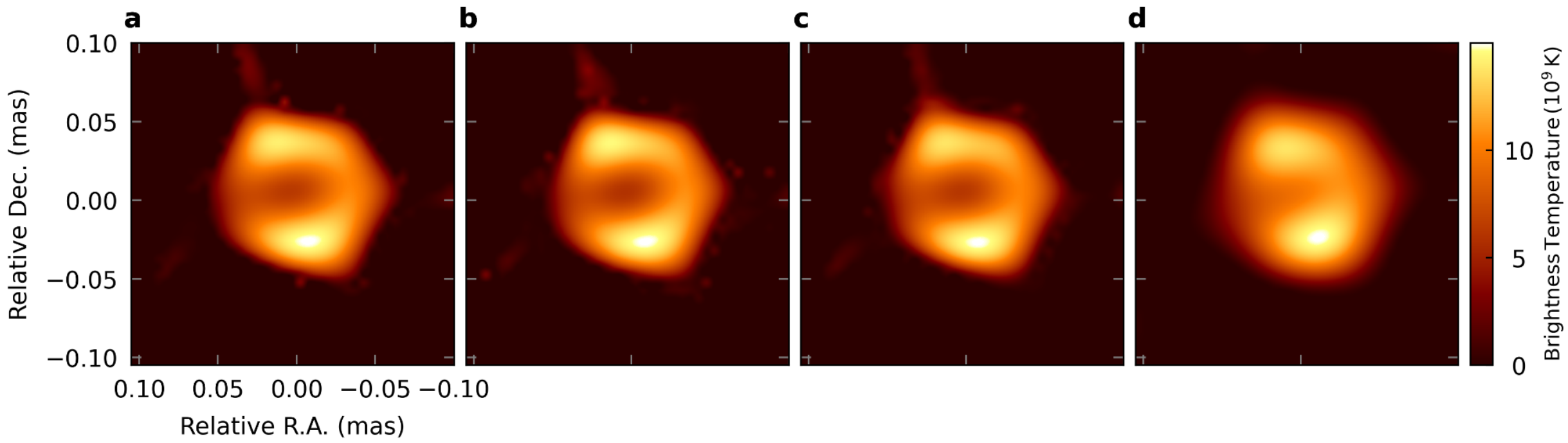

**Figure S6| Dependence of the ring-like structure on SMILI regularizers.** Images are reconstructed with the hyper parameter (weight) on the total squared variation (TSV) varying from 0 (**a**) to 10 (**b**) to 100 (**c**) to 1000 (**d**). All other imaging parameters are fixed (total flux = 0.7 Jy, weighted-L1 = 1, TV = 0, MEM = 0, and Gaussian FWHM = 50 μas for convolution of the pre-processed image used as prior for wL1, see Supplementary section 3 for details).

## Synthetic data imaging tests

Although all images of the core show a faint underlying ring-like structure, we noticed that in the CLEAN images, the dominance of the two blobs appear to be more significant than in some of the SMILI images, which show more azimuthally uniform ring-like structures. We postulate that this difference originates from: (i) the limited $(u,v)$-coverage of our data, and (ii) the fact that the RML technique is better suited for super-resolution imaging than CLEAN. To demonstrate this, we have created synthetic data sets assuming ring models with different diameters from 52 to 120 μas using the eht-imaging library[26]. For the ring, we adopted a Gaussian brightness distribution with a FWHM of 20 μas and a total flux of 0.5 Jy. We added thermal noise to the data based on the visibility weights and assumed that there are no additional gain errors in order to check the impact of the limited $(u,v)$-coverage and to compare the super-resolution imaging performance between the CLEAN and the RML techniques. The synthetic data sets were generated for two different $(u,v)$-coverages. One is based on the real $(u,v)$-coverage, 10-second averaged data and the other is generated by adding artificial baselines to VLBA MK and KVN Yonsei. These artificial baselines were assumed to have constant visibility weights similar to the GBT-KP baseline. We present the (*u,v*)-coverage of these synthetic data in Figure S7. The artificial baselines provide high sensitivity very long baselines along the east-west direction, which is missing in our data. Thus, this data will help us to understand the importance of the long EW baselines in the image reconstruction.

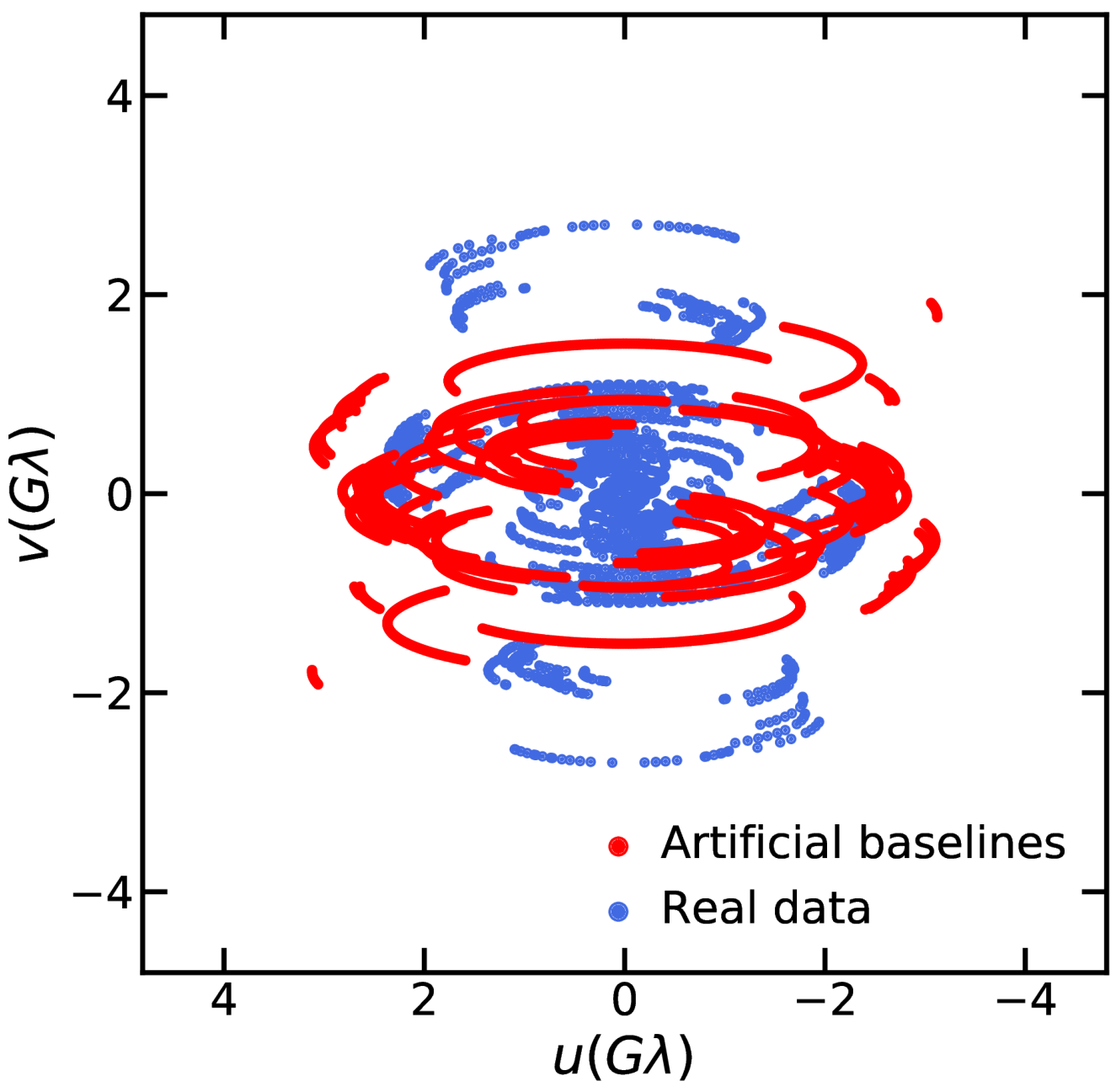

**Figure S7 | (*u,v*)-coverage of the simulated data.** Simulated (*u,v*)-coverage for the 2018 observations of M87 (blue) and for the additional baselines to VLBA MK in Hawaii, USA and KVN Yonsei in South Korea (red), added for the purpose of investigating the effects of (*u,v*)-coverage on the results.

We present the reconstructed images with SMILI and CLEAN in Figure S8. In general, the reconstructed images from SMILI show a smoother structure than the CLEAN images for both synthetic data sets. Interestingly, the images for the data based on the real data (*u,v*)-coverage with ring diameters of 56-68 μas (similar to the observed diameters) show bright and symmetric two arc-like structures at a position angle of about 25°. This structure is quite similar to the SMILI images from the real data, although the dominance of the arc-like structures becomes less pronounced when the diameters become larger. The CLEAN images for the same synthetic data set present two hot-spots on top of the ring at a similar position angle, which is also similar to the CLEAN images from the real data. However, these features disappear or become much weaker when the synthetic data set with artificial baselines was used for both SMILI and CLEAN reconstructions. The above results indicate that the detailed features on top of the underlying ring-like structure can be affected significantly by (*u,v*)-coverage, with the caveat that residual calibration uncertainties with real data and different assumptions on the sensitivity of the artificial baselines may further affect the reconstruction of the ring properties (e.g., ring width).

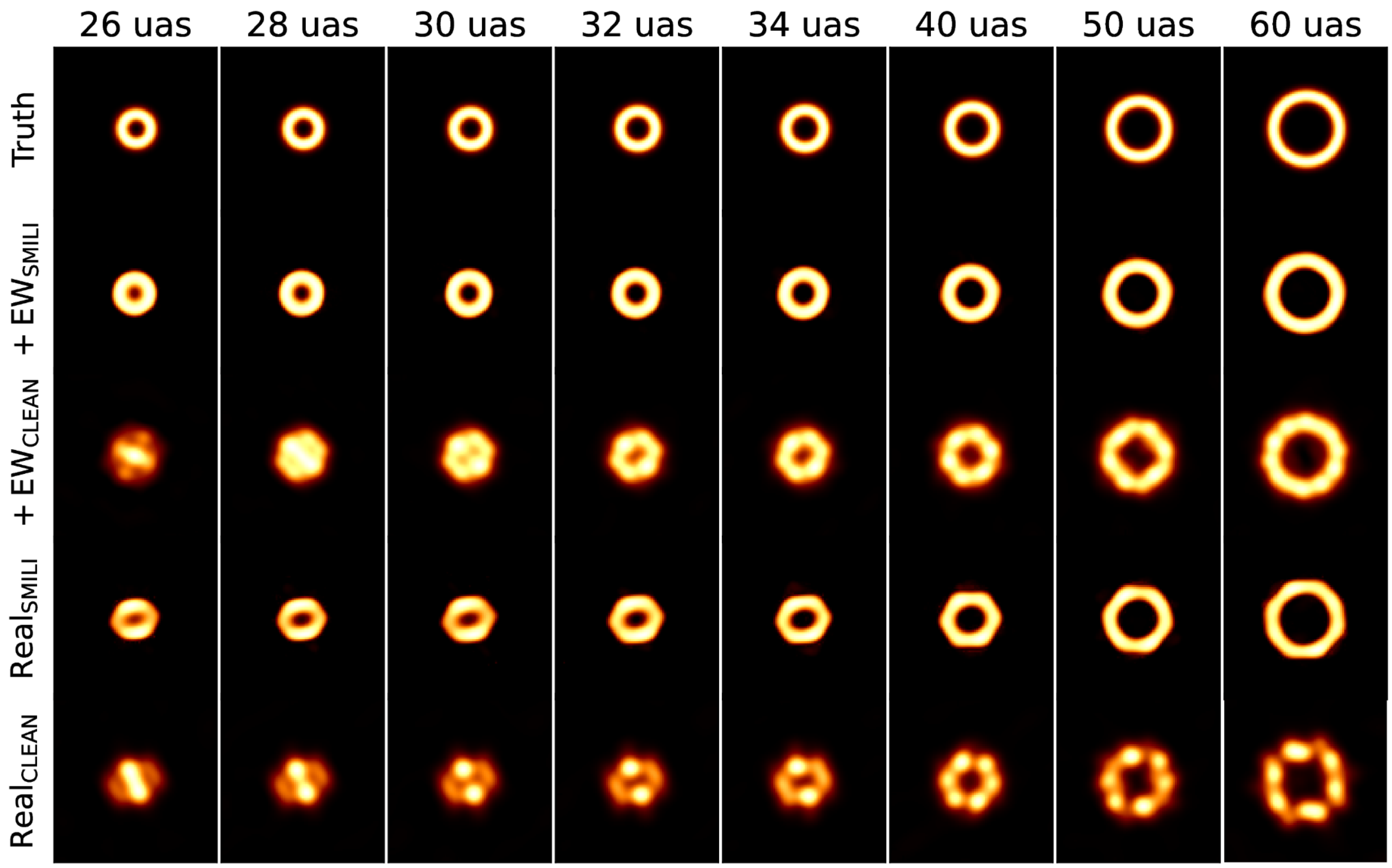

**Figure S8 | Image reconstruction with SMILI and CLEAN using the synthetic data.** The results for different ring diameters are presented in the different columns. The ground truth images used for the synthetic data generation are shown in the top row. The second and third rows from the top show the results from SMILI and CLEAN imaging based on the (*u,v*)-coverage with artificially added east-west baselines, respectively, while the fourth and fifth rows show images based on the (*u,v*)-coverage of the real data. The CLEAN images are displayed after convolution with a circular beam size of FWHM= 30 μas.

**5. Feature extraction for the spatially resolved core**

We derive the characteristic properties of the ring-like structure using the optimal set of SMILI images. Due to the limited (*u,v*)-coverage of our observations (see Figure S7), we restrict our analysis to circular ring fitting. The method we used here is similar to the one used for the analysis of the EHT observations of M87 at 1.3 mm (ref.[6]).

We started by identifying a tentative origin of the ring-like structure for each reconstructed image, which is close to the centre of the central depression in the brightness distribution. Using this tentative position as the origin, we unwrapped the image azimuthally and obtained 100 radial profiles in 3.6-degree steps between 0 and 360°. For each azimuthal angle, we identified the brightness peak in the radial profile. We then fit a circle to the brightness peak positions across the various azimuthal angle profiles. Since the tentative origin does not necessarily coincide with the fitted ring origin, we also fit the ring central position (x0, y0) in addition to the ring radius. The ring parameters are estimated by minimising the standard deviation of the ring radius. Based on the new origin estimate, we perform the same procedure and refine the ring parameters. We iterate this process until the solution converges. There are some radial profiles where the brightness monotonically decreases with radius (see, e.g., Figure S6c). In this case, we discarded these profiles in the ring fitting. After determining the origin of the circular ring, we derive the ring width ($w$). We first derived cross-sections of the core structure which pass through the origin of the ring at azimuthal angles between 0 and 180°. We then fitted a double Gaussian to characterise each cross section. The width is determined by the mean of the FWHM of the Gaussian components at all position angles.

In Figure S9, we show the derived mean diameter and width of the ring-like structure for all of the optimal set of images. For both the diameter and width, we adopt the median as their fiducial value and the 95% highest density interval to characterise their corresponding uncertainty. We found that the diameter is $64^{+4}_{-8}$ μas and the width is $38^{+7}_{-17}$ μas. We note that the width should only be considered as an upper limit due to the finite array resolution[9] and the results from this optimal set of images do not formally correspond to a posterior distribution. Nonetheless, these images cover a reasonably wide space of imaging parameters and are compatible with the data.

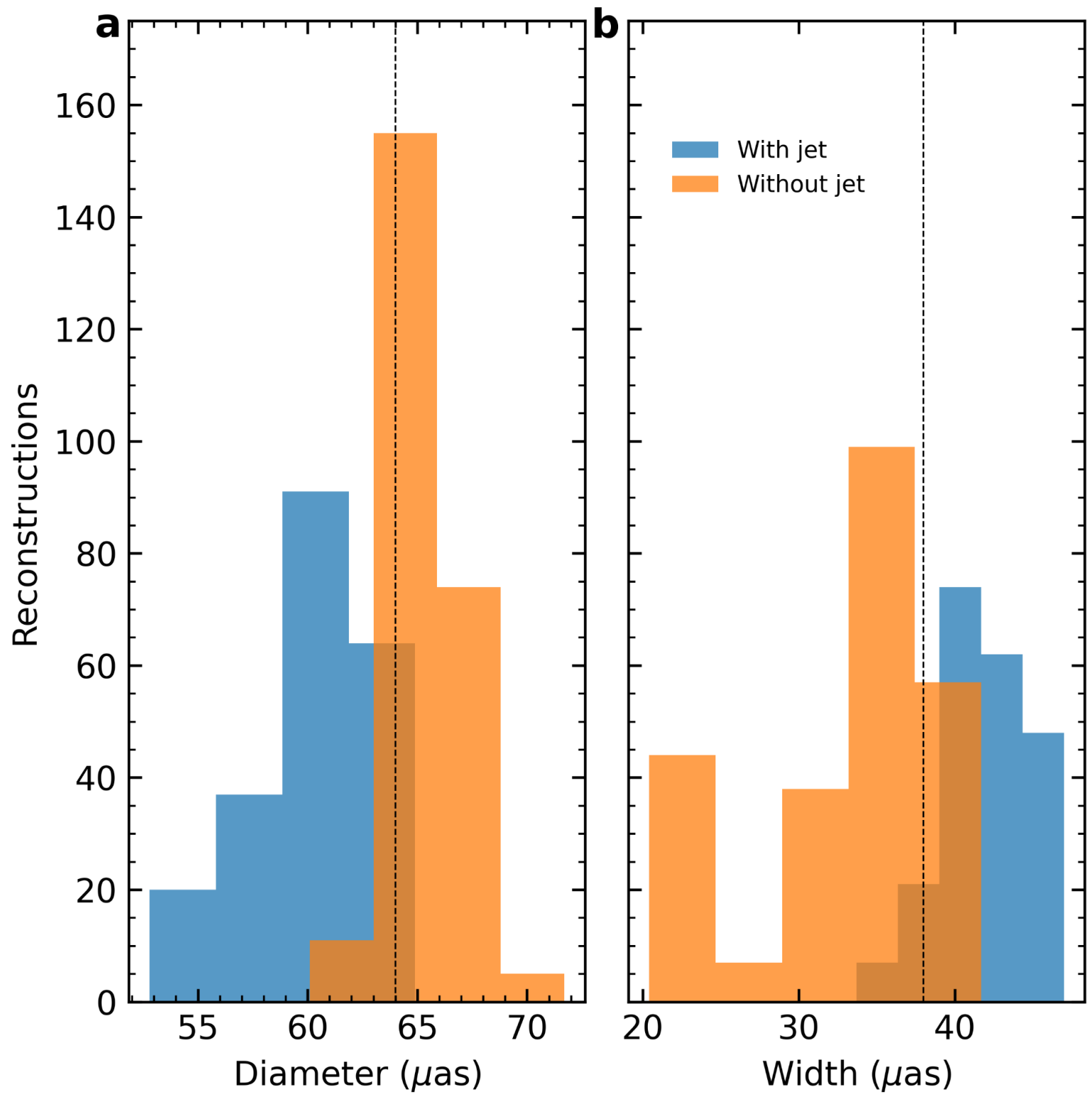

**Figure S9 | Ring fitting results of the optimal set of images of the core. a,** histogram of the mean ring diameter. **b,** histogram of the ring width. In each plot, the orange colour denotes the optimal set of images using the jet-subtracted data while the blue colour denotes the optimal set of images using the full data. The vertical dashed lines mark the median diameter and width, respectively. Note that the distribution originates from different imaging parameters and realisation of self-calibration. For both the diameter and width, the median values of the two methods (i.e., with or without jet subtraction) agree with the adopted fiducial values within one pixel (5 μas) of the full data imaging.

## 6. Visibility domain model fitting

In another approach (and independently from the direct imaging and image plane analysis), we study the visibility data directly in order to investigate whether a specific morphological type of the structure is favoured by the data. For this purpose, the visibility amplitudes have been model fitted by a set of five circularly symmetric patterns[46,47]: 1) circular Gaussian, 2) a uniformly bright disk, 3) a uniformly filled sphere, 4) an infinitesimally thin ring (hereafter "thin ring"), and 5) a uniformly bright ring of finite thickness (hereafter "thick ring"). The first four models are described by two parameters: the total flux density, $V_0$, and the diameter $d$ (proxied by the full width at half maximum, FWHM, for the Gaussian model). The thick ring model has the ring width, $w$, as an additional parameter.

This fitting was applied to both the full self-calibrated dataset and to the jet-subtracted dataset. In each case, the data were averaged to scan length (420 seconds). The data weights were determined by the correlator weights, $\eta_i$, of the visibilities, and the respective errors of visibility amplitudes were set by $\sigma_i = \eta_i^{-\frac{1}{2}}$. The visibility averaging was done in Difmap which is known to bias the r.m.s. noise in the averaged data. Comparisons of the r.m.s. noise in the original and averaged data indicated noise reductions by factors of 2.28 and 2.97 for the full and jet-subtracted datasets, respectively. These factors were applied retroactively to the visibility errors of the averaged data in order to maintain the r.m.s. noise of the original data.

Because of the circular symmetry of the fitted models, it sufficed to apply the fitting solely to the radial distances, $q$ ($\equiv\sqrt{u^2 + v^2}$), of the visibility measurements in the $uv$-plane. Figures S10 and S11 show the resulting fits, and the respective parameters of the fits (total flux density, $V_0$, and diameter, $d$, for all models and width, $w$, for the ring with finite thickness) are summarised in Tables S1 and S2, together with the reduced $\chi^2$ goodness-of-fit parameters. We note that the formal errors on the fitted parameters should be interpreted with care given the simplicity of the model.

It is clear from Tables S1 and S2 that, for both datasets, the thick ring fit achieves the best results, as reflected by the respective reduced $\chi^2$ parameters. The fitted diameters of the thick ring (66.8 μas and 66.0 μas) are similar to the ring diameter of 64 μas obtained from the RML imaging of the data. Both are about a factor of 1.6 larger than the ring diameter obtained from the EHT observations[3], indicating that opacity in the accretion flow may play a role in the differences of the ring diameters as measured at 3.5 and 1.3 mm (see also the discussion in the main text).

The visibility amplitudes for both the full and the jet-subtracted datasets indicate that the first zero amplitude (first null) point required by the non-Gaussian models should be most likely located at $q_{\rm null}$ ~ 2.3 Gλ, with the potential radial range for this point extending between ~ 2.1 Gλ and ~ 2.6 Gλ (Figures S10 and S11).

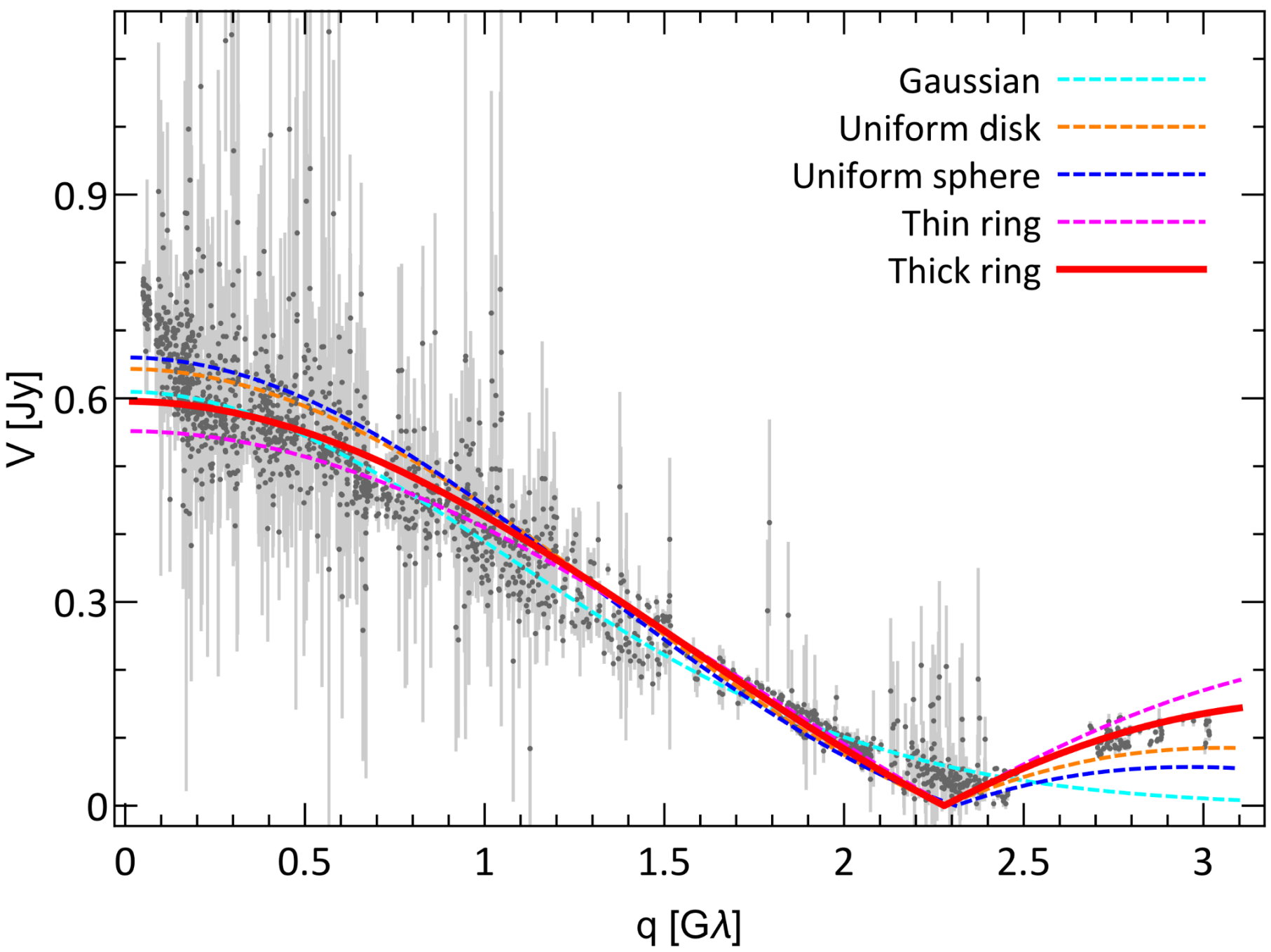

**Figure S10 | Model fits by circularly symmetric patterns applied to the full, self-calibrated dataset.** Effect of the extended jet emission is visible at short (*u,v*)-distances (i.e., small *q*). The extended jet contribution becomes less important at (*u,v*)-distance $q \geq 1.5$ Gλ, where the visibility function fully reflects the most compact source structure. Error bars are 1$\sigma$. Parameters of the respective model fits are listed in Supplementary Table S1.

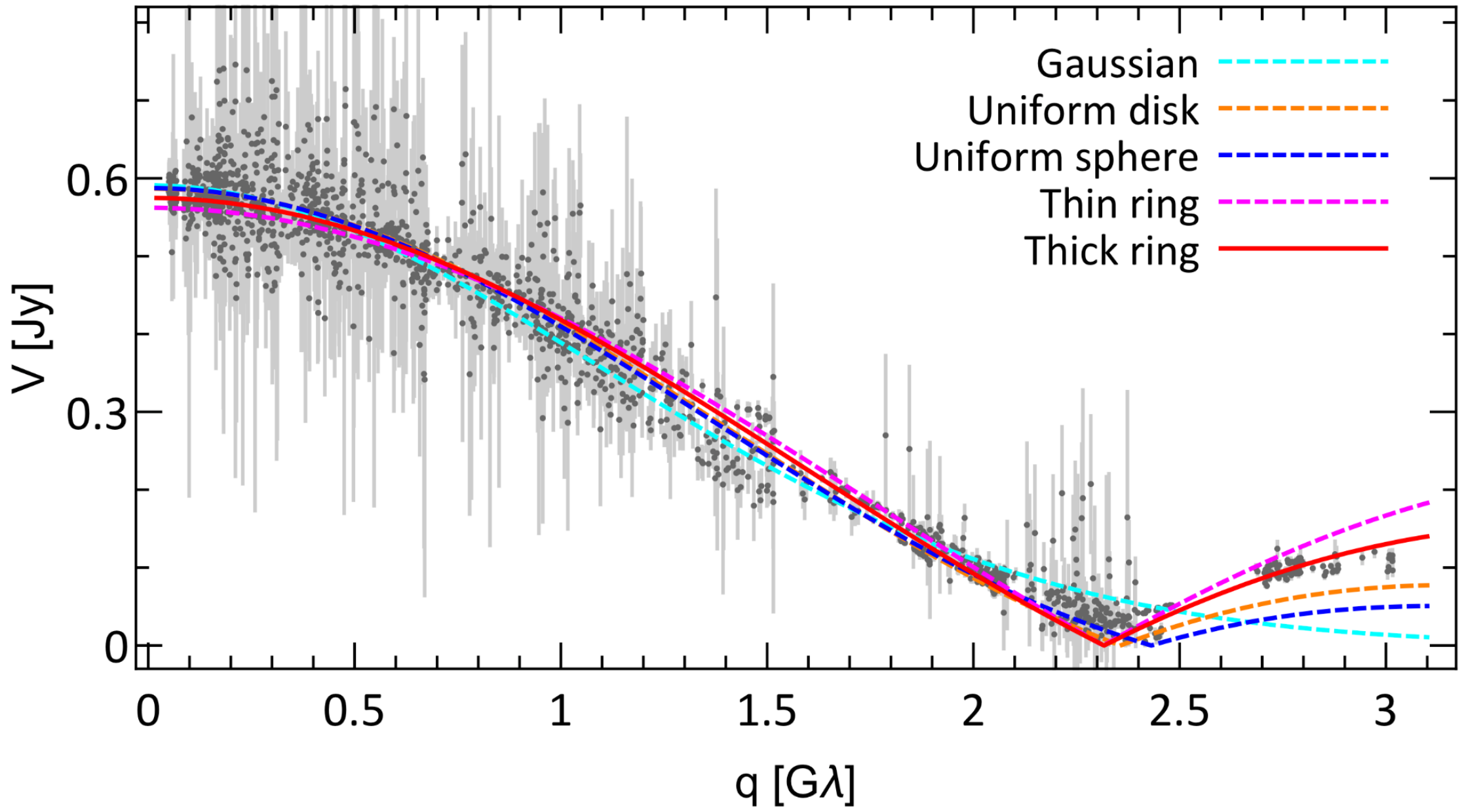

**Figure 11 | Model fits by circularly symmetric patterns applied to the jet-subtracted dataset.** The data have been self-calibrated as described in Supplementary section 3 before jet-subtraction. Error bars are 1$\sigma$. Since the contribution of the extended jet emission (mostly visible at short $(u,v)$-distance, i.e., small $q$) has been removed, this data set is better suited to discriminate between different compact emission models. The parameters of the corresponding models are summarised in Supplementary Table S2.

**Table S1 | Parameters of model fits applied to the full, self-calibrated dataset.**

| Model | $V_0$ [mJy] | $d$ [μas] | $w$ [μas] | $\chi^2$ |
|---|---|---|---|---|
| Circular Gaussian | 609.6 ± 4.3 | 73.3 ± 0.2 | | 9.0 |
| Uniform disk | 643.4 ± 2.8 | 110.3 ± 0.1 | | 5.9 |
| Uniform sphere | 660.2 ± 3.9 | 127.7 ± 0.2 | | 10.1 |
| Thin ring | 551.7 ± 2.0 | 69.0 ± 0.1 | | 4.7 |
| Thick ring | 595.6 ± 2.5 | 66.8 ± 0.1 | 29.8 ±0.7 | 3.5 |

**Table S2 | Parameters of model fits applied to the jet-subtracted dataset.**

| Model | $V_0$ [mJy] | d [μas] | $w$ [μas] | $\chi^2$ |
|---|---|---|---|---|
| Circular Gaussian | 591.0 ± 2.1 | 70.7 ± 0.2 | | 5.21 |
| Uniform disk | 586.8 ± 1.4 | 106.8 ± 0.2 | | 4.67 |
| Uniform sphere | 587.5 ± 1.8 | 121.3 ± 0.3 | | 7.62 |
| Thin ring | 562.1 ± 1.3 | 68.2 ± 0.1 | | 2.10 |
| Thick ring | 574.8 ± 1.3 | 66.0±0.7 | 27.6 ± 0.6 | 1.85 |

To evaluate the robustness of the fitting result, we fit the data with the thick ring model for outer ring radii, $r_{out}$, ranging between 34.5 and 54.5 μas. The respective models depend only on the total flux, $V_0$, and the ring width, $w$. The inner, $r_{in}$, and mean, $r_c$, ring radii can be readily calculated from the fits. These fits approximately probed the 2.1–2.6 Gλ range of $q_{null}$. The relationship between the ring radii and the $\chi^2$ goodness-of-fit parameter are plotted in Figure S12 and a subset of the resulting models is shown in Figure S13.

The best fit obtained for this set of models is model ***e)*** in Figure S13. It is very similar to the best fit generic thick ring model shown in Figure S11 and Table S2. Evolution of the models presented in Figures S12 and S13 shows that the outer radius and width of the ring are correlated and that this correlation leads the model to approach the thin ring at $r_{out}$ = 34.5 μas (model ***a)***) and the uniform disk at $r_{out}$ = 54.5 μas (model ***i)***), hence covering the entire range of plausible model representations. Using the Kolmogorov-Smirnov test on the residuals of the fits[48], the confidence intervals can be estimated (Table S3).

It is apparent that all confidence intervals except for the formal 1$\sigma$ errors are strongly constrained by the result of the thin ring fit for which all three radii approach the value of 34.5 μas. In addition, we note the formal best-fit size of the inner depression as described by $r_{in}$ is about 1.7 times larger than the inner radius of ~ 13 μas measured from the EHT data[4]. Notably, the lower boundary of the inner radius, as constrained by the GMVA data, agrees with the respective EHT value up to the 90% confidence level. Altogether, the analysis presented above suggests the presence of a ring-like structure in the centre of M87 which has a mean diameter of ~ 67 μas and a width of ~ 22 μas.

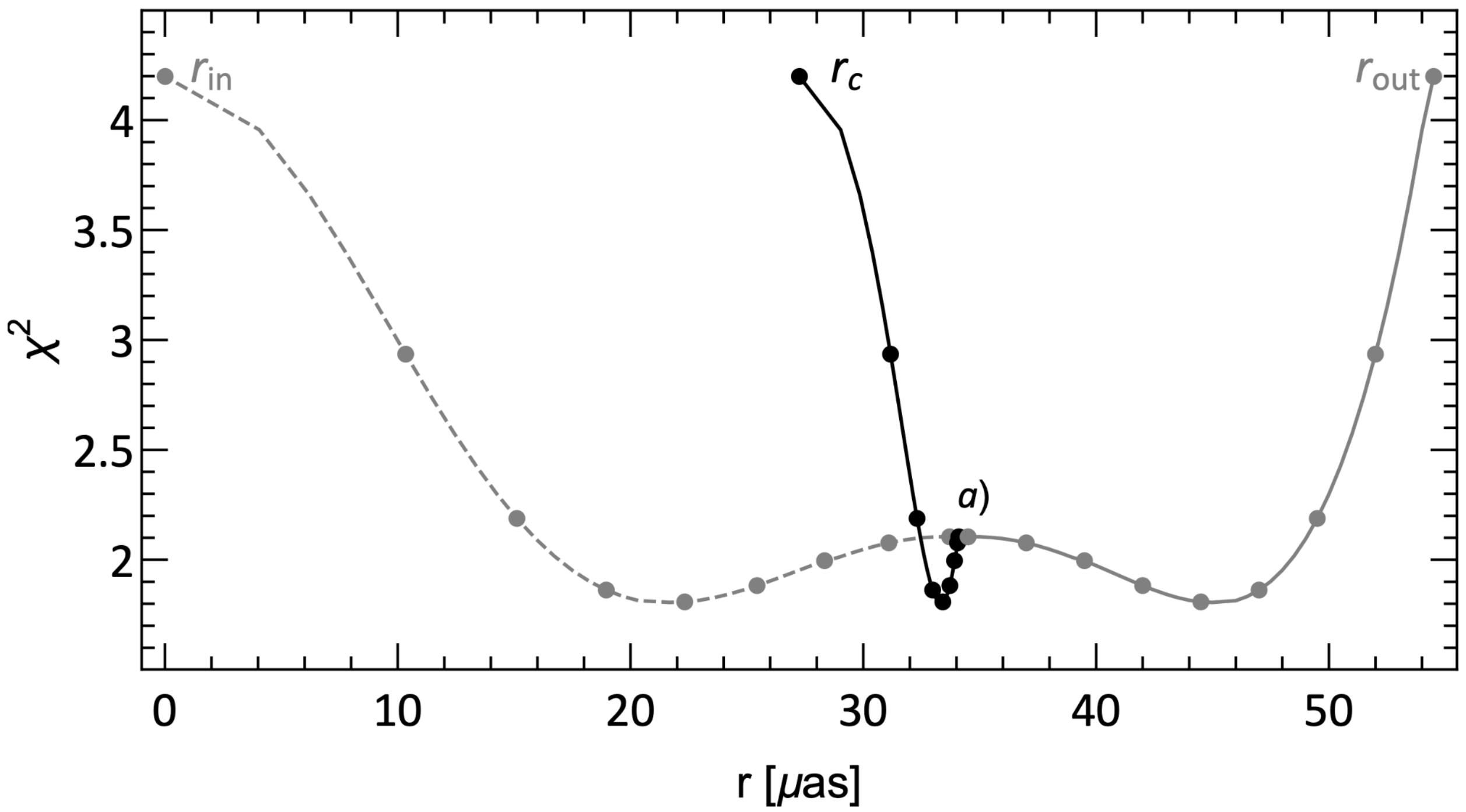

**Figure S12 | Evolution of the $\chi^2$ goodness-of-fit parameter.** Reduced $\chi^2$ is obtained from fitting the jet-subtracted dataset with the thick ring models calculated for outer ring radii, $r_{out}$, ranging from 34.5 μas to 54.5 μas. The respective mean, $r_c$ (solid line), and inner, $r_{in}$ (dashed line), ring radii are also shown. Dots indicate the fits by the subset of the models shown in Figure S13, with the symbol *a)* indicating the first (top left) model in that figure.

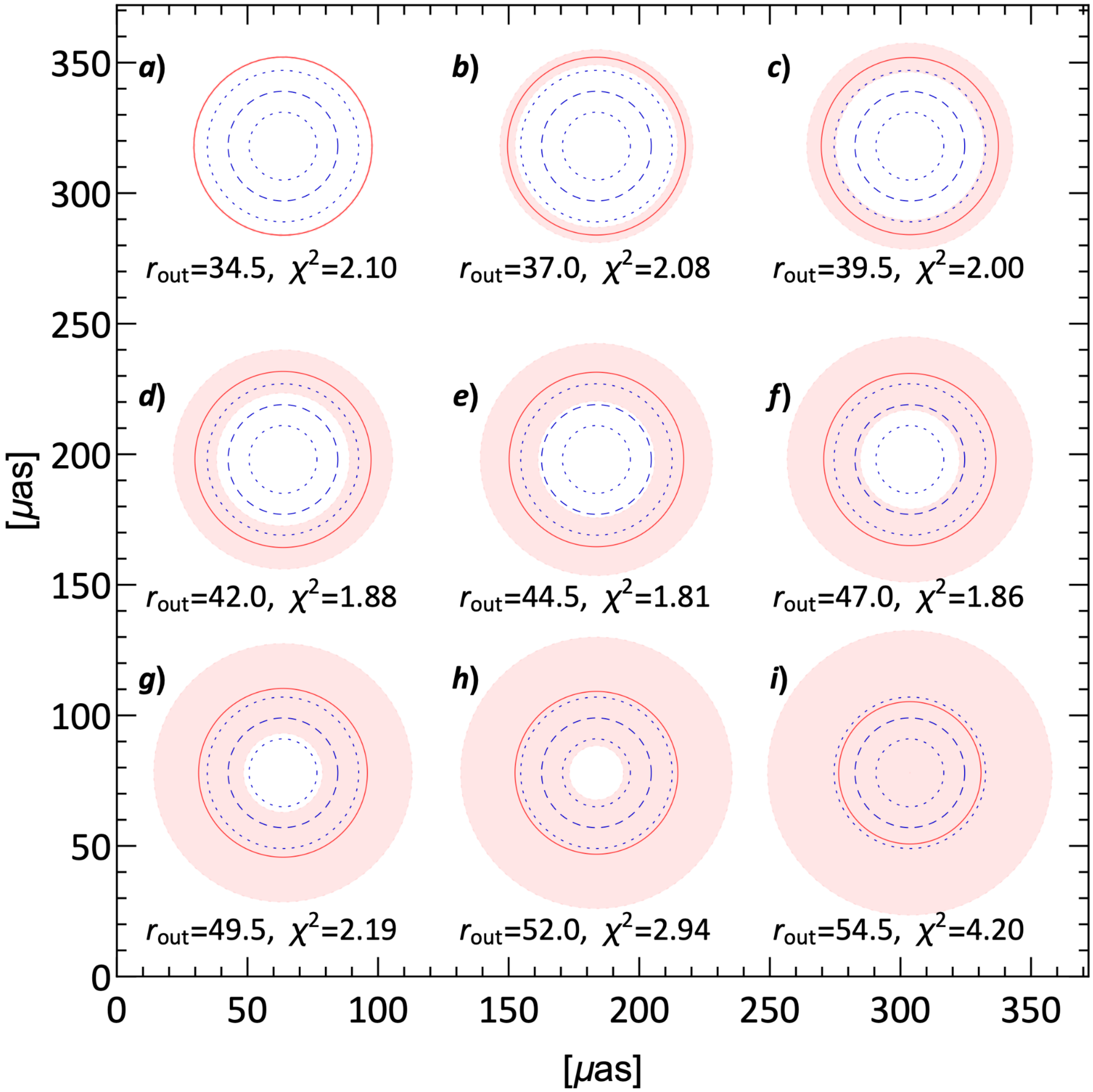

**Figure S13 | Subset of thick ring models obtained for different values of the outer ring radius, $r_{out}$.** For each model, the mean ring radius, $r_c$, is shown by the red circle and the area between the inner, $r_{in}$, and outer, $r_{out}$, ring radii is shaded in pink. For the purpose of comparison, the respective ring parameters obtained from the EHT observations[4] are shown in blue colour (dashed circle for the mean ring radius, and dotted lines for the respective inner and outer ring radii). Models shown in panels ***a)***, ***e)***, and ***i)*** approximately correspond to the thin ring, the thick ring, and the uniform disk models presented in Figure S11 and Table S2.

**Table S3. Best fit value and confidence interval for the thick ring model.**

| Ring radius | Best fit value | Confidence interval | | |
|---|---|---|---|---|
| | | c.l. 90.0% | c.l. 95.4% ($2\sigma$) | c.l. 99.7% ($3\sigma$) |
| inner, $r_{in}$ [μas] | 22.3 (+7.0, -7.1) | (12.3, 34.5) | (11.2, 34.5) | (9.2, 34.5) |
| mean, $r_c$ [μas] | 33.4 (+0.6, -1.1) | (31.6, 34.5) | (31.4, 34.5) | (30.8, 34.5) |
| outer, $r_{out}$ [μas] | 44.5 (+4.9, -5.9) | (34.5, 51.0) | (34.5, 51.6) | (34.5, 52.4) |

## 7. Summary of the properties of the ring-like structure based on image and visibility domain analysis

The derived diameter from the image domain analysis $64^{+4}_{-8}$ μas is in good agreement with that from the visibility domain fitting ($67^{+2}_{-4}$ μas, errors are 2 σ, see Table S3). Here we adopt $64^{+4}_{-8}$ μas as the fiducial value for the diameter of the ring-like structure, which is ~ 22 μas (i.e., 50%) larger than the 1.3 mm ring measured with EHT observations[4]. Although all images and valid models from the visibility domain fitting show a central depression, the fitted widths have less consistency between the two methods, with the best-fit width from the visibility fitting being close to the lower bound of the widths extracted from the imaging analysis (both are ~ 20 μas). For the image domain model fitting, it has been shown that the ring diameter can be biassed downward due to the finite resolution and the non-negligible width with respect to the ring diameter (ref.[6], Appendix G). In Figure S14, we show the fractional width (i.e., width/diameter) as a function of the mean diameter for the optimal set of SMILI images, which indicate the existence of such a correlation as expected.

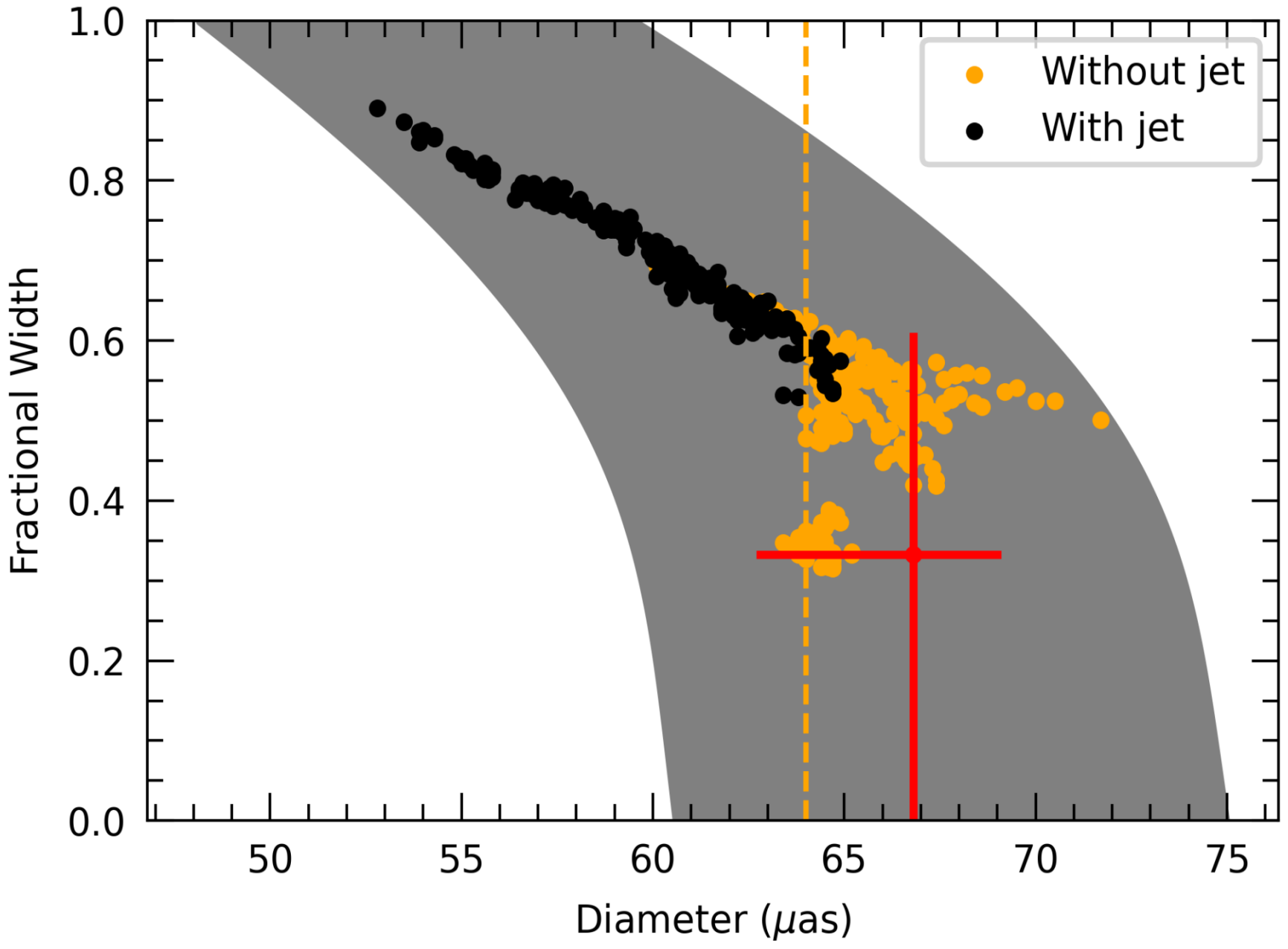

**Figure S14 | Correlation of the ring width and diameter.** Fractional widths are plotted as a function of the mean diameter for the optimal set of SMILI images. The vertical dashed line marks the median value of 64 μas. The red cross denotes the best fit thick ring model from the visibility domain analysis, with 2σ confidence intervals for the derived diameter and fractional width of the ring. The shaded area denotes the expected anticorrelation between diameter and fractional width for the observed minimum of the visibility amplitudes between 2.1 and 2.6 Gλ for a geometric crescent model (see Section 7.2 in ref.[4] for more details).

## 8. Measuring the jet collimation profile

We constructed the jet collimation profile based on the mean CLEAN image that was convolved with an 80 μas circular beam for the outer part of the jet (0.1-0.65 mas) and a 37 μas circular beam for the inner part (0.02-0.1 mas). These beam sizes correspond to the major and minor axis of the Gaussian CLEAN restoring beam. For both cases, we rotated the image by 23° clockwise such that the jet features at ~ 0.2 mas from the reference point (i.e., the centre of the ring-like structure) in the northern rim and ~ 0.4 mas from the reference point in the southern rim has the same angular distance to the horizontal axis. This assumes an overall jet position angle of -67° and is consistent with previous VLBI observations at 86 GHz[5,28].

We made slices transverse to the jet every 25 μas for the inner part from the reference point and every 50 μas for the downstream jet. For all slices, two or (mostly) three well-resolved Gaussians are needed to fit the sampled intensity profile, allowing us to accurately determine the jet width for each slice. We adopted the distance between the outer edges of the FWHMs of the two outermost Gaussians as the jet width. Uncertainties in the measured jet width take into account a minimum uncertainty determined by our resolution limit (here we adopt ⅕ beam size)[49] or ⅕ width of the fitted Gaussian (whichever is larger) and fitting uncertainties. The jet width profile is shown in Fig. 3 in the main text.

**9. Theoretical simulations of the core structure**

**Property of the general relativistic magnetohydrodynamic (GRMHD) simulation data** We performed an axisymmetric 2D GRMHD numerical simulation for magnetised plasma around a rotating black hole by using the public code HARM[50]. In previous EHT modelling works, prograde black hole spin parameters up to 0.94 were considered[12,51]. In addition, spin parameters within the range of 0.5 - 0.99 can reproduce the observed large scale jet width[19]. Therefore, we consider a dimensionless black hole spin parameter of 0.9 for our reference simulation run. The radiation feedback is not essential for the dynamics of a RIAF and is therefore ignored in the GRMHD simulation. Starting with a magnetised torus, the initial setup and boundary conditions are similar to previous studies[12]. The GRMHD simulation can be scaled to different black hole mass values and different normalisations of the plasma density. A snapshot is taken from the simulation data during the magnetically arrested disk phase (the dimensionless magnetic flux $\varphi \sim 60$).

**General relativistic radiative transfer** The model image can be computed by postprocessing the numerical simulation data, by applying general relativistic ray tracing (GRRT), given that the scattering of photons in the radio band is not essential. Taking into account the relativistic effects, in the GRRT, observed rays are traced backwards in time and the emission along the null geodesics is integrated. The parameters related to the GRRT computation are summarised below.

For modelling the core emission and image structure of M87, we consider a black hole of $6.5 \times 10^9$ solar masses at a distance of 16.8 Mpc, and assume the inclination angle between the black hole rotational axis and the distant observer is 163° (corresponding to a standard inclination of 17° between the approaching jet and the line of sight). To explore the relative importance of the jet and accretion flow, we consider two models with synchrotron emission from different electron energy distributions and different regions.

For the non-thermal synchrotron model, synchrotron emission is solely from electrons with a power-law energy distribution in the jet region, defined by $b^2/\rho c^2>1$ ($b$ is magnetic field strength, $\rho$ is the mass density and $c$ the speed of light). The power-law index in electron energy is assumed to be constant $p$=3.5 between the cutoffs at low- and high-energy, with the low-energy cutoff $\gamma_{min}$=50. Assuming the internal energy of the non-thermal electrons is proportional to the magnetic energy in the jet region, we follow the formula and procedure described in previous studies[52] for modelling the M87 jet emission. For the thermal synchrotron model, synchrotron emission is solely from electrons with a relativistic Maxwellian energy distribution in the accretion flow region, defined by $b^2/\rho c^2<1$. A constant ratio ($T_i/T_e$ = 4) between ion temperature ($T_i$) and electron temperature ($T_e$) is applied.

The corresponding spectrum of each model is presented in Figure S15. The normalisation of the plasma density is determined by the observed core flux at 1.3 mm (the black square data point). For both models, the system is still optically thick at 86 GHz (3.5 mm), but becomes optically thin at $\gtrsim$230 GHz ($\lesssim$1.3 mm). The model images at these two frequencies are shown in Fig. 2 in the main section of this paper.

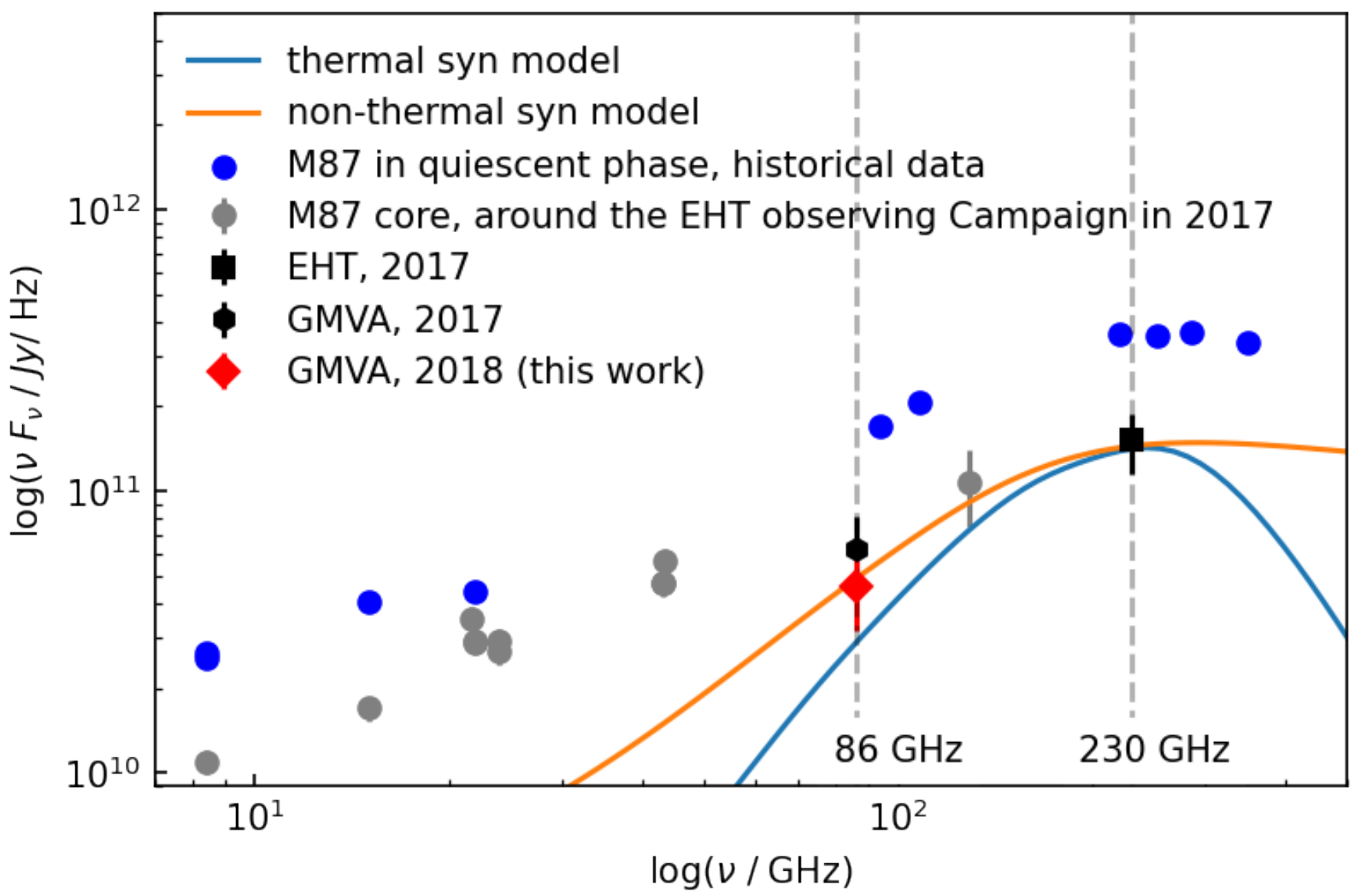

**Figure S15 | Spectral energy distribution (SED) of the model images.** The blue curve corresponds to the model of thermal synchrotron emission from the accretion flow region. The orange curve corresponds to the model of non-thermal synchrotron emission from the jet region. The observed 3.5 mm ring-like structure can be explained by the thermal synchrotron model, but not by the non-thermal synchrotron model (see Fig. 2 of the main text for the corresponding model images at 3.5 mm and 1.3 mm for both models). The SED (in quiescent phase) from a ~ 0.4 arcsecond aperture radius that include arcsecond-scale jet emission[53] (blue data points) and the VLBI flux densities obtained during the 2017 EHT campaign at 1.3 mm and at longer wavelengths[54] are shown as black and grey symbols, respectively. Error bars represent 1$\sigma$.

## Supplementary References